\documentclass[12pt]{article}
\pdfoutput=1

\usepackage[a4paper,text={16.8cm,22.4cm}]{geometry}
\usepackage{amsmath,amsfonts,braket,slashed,amssymb,bm,psfrag,graphicx,color,colortbl,dsfont,euscript,ctable,bbm}
\usepackage[small,labelfont=bf]{caption}
\RequirePackage[sort&compress,square,comma,numbers]{natbib}
\allowdisplaybreaks
\newcommand{\spac}{{\hspace{0.3mm}}}
\definecolor{LightGray}{rgb}{0.85,0.85,0.85}
\definecolor{VeryLightGray}{rgb}{0.95,0.95,0.95}

\begin{document}

\begin{titlepage}

\begin{flushright}
MITP-23-039\\
August 2, 2023
\end{flushright}

\vspace{3mm}
\begin{center}
\Large\bf
ALP\,\hspace{-0.3mm}--\,LEFT Interference and the Muon $\bm{(g-2)}$
\end{center}

\vspace{-1mm}
\begin{center}
Anne Mareike Galda$^a$ and Matthias Neubert$^{a,b}$\\
\vspace{0.7cm} 
{\sl ${}^a$PRISMA$^+$ Cluster of Excellence \& Mainz Institute for Theoretical Physics\\
Johannes Gutenberg University, 55099 Mainz, Germany\\[2mm]
${}^b$Department of Physics \& LEPP, Cornell University, Ithaca, NY 14853, U.S.A.\\[2mm]
}
\end{center}

\vspace{1.5cm}

\begin{abstract}
The low-energy effective field theory (LEFT) provides the appropriate framework to describe particle interactions below the scale of electroweak symmetry breaking, $\mu_w\sim v$. By matching the Standard Model onto the LEFT, non-zero Wilson coefficients of higher-dimensional operators are generated, suppressed by the corresponding power of $1/v$. An axion or axion-like particle (ALP) with mass $m_a\ll\mu_w$ that interacts with the Standard Model via classically shift-invariant dimension-five operators would also contribute to the LEFT Wilson coefficients, since it can appear as a virtual particle in divergent Green's functions and thus has an impact on the renormalization of the LEFT operators. We present the full set of one-loop ALP-induced source terms modifying the renormalization-group evolution equations of the LEFT Wilson coefficients up to dimension-six order. Our framework allows for model-independent ALP searches at low energies from current bounds on LEFT Wilson coefficients. As a concrete application, we present an improved prediction for ALP effects on the anomalous magnetic moment of the muon. 
\end{abstract}

\vspace{0.8cm}

\end{titlepage}

\tableofcontents

\section{Introduction}

Due to their potential to explain the absence of CP violation in the strong interactions and thereby solve the so-called ``strong CP problem'', axions are particularly well-motivated candidates for new degrees of freedom extending the Standard Model (SM) of particle physics \cite{Peccei:1977hh,Weinberg:1977ma,Wilczek:1977pj}. They are pseudo Nambu--Goldstone bosons from a broken global $U(1)_{\mathrm{PC}}$ Peccei--Quinn symmetry. Generalizing this concept, new light degrees of freedom arising from a spontaneously broken global symmetry are commonly referred to as axion-like particles. They appear for instance in solutions suggested for the flavor problem \cite{Ema:2016ops,Calibbi:2016hwq}. In the following, axions and axion-like particles will be commonly referred to as ALPs. Due to a classical shift symmetry, possible couplings to the SM particles are of dimension-five order and higher, yielding a natural suppression of these operators by powers of the scale of global symmetry breaking $\Lambda$. Search strategies for ALPs use a large variety of cosmological observations \cite{Cadamuro:2011fd,Millea:2015qra}, astrophysical measurements \cite{Payez:2014xsa,Jaeckel:2017tud}, and collider probes \cite{Mimasu:2014nea,Jaeckel:2015jla,Knapen:2016moh,Brivio:2017ije,Bauer:2017ris,Bauer:2018uxu}, as well as precision studies of flavor-violating transitions in the quark and lepton sectors  
\cite{Batell:2009jf,Freytsis:2009ct,Dolan:2014ska,MartinCamalich:2020dfe,Bauer:2019gfk,Bauer:2021mvw}. Most of these analyses are strongly model dependent. For example, direct searches for ALPs at particle colliders must make an assumption about the process in which the ALP is produced (e.g.\ in the decay of a Higgs boson, a $Z$ boson, or in weak decays of kaons or $B$ mesons), about its lifetime, and the way in which it decays, which may involve a decay outside of the detector. In the derivation of astrophysical bounds on the ALP--photon coupling from searches for ALPs produced in supernovae or in the sun, it makes an important difference whether or not a vanishing ALP--electron coupling is assumed. In the majority of the existing analyses, it was assumed that only a single ALP coupling is non-zero at the scale of Peccei--Quinn symmetry breaking -- an assumption which is not valid in even the classic KSVZ \cite{Kim:1979if,Shifman:1979if} and DFSZ \cite{Dine:1981rt,Zhitnitsky:1980tq} models for the QCD axion.

Indirect searches for ALP couplings, which look for subtle quantum effects in precision measurements arising from virtual ALP exchange, offer a promising alternative strategy to look for ALPs and their interactions with SM particles in a more model-independent way. In recent work, we have described the phenomenon of the ``ALP\,--\,SMEFT interference'' \cite{Galda:2021hbr}, which refers to the fact that ultraviolet (UV) divergences in loop diagrams with virtual ALP exchange require as counterterms operators built out of SM fields only, and this affects the renormalization-group (RG) evolution of the operators in the Standard Model effective field theory (SMEFT). Indeed, we have shown that the presence of an ALP with non-zero couplings to the SM induces inhomogeneous source terms in the RG evolution equations of the dimension-six Wilson coefficients of almost all SMEFT operators in the Warsaw basis \cite{Grzadkowski:2010es}, starting at one-loop order. In the present paper, we extend this concept to the effective operators present below the scale of electroweak symmetry breaking, $\mu_w\sim v$ (with $v\approx 246.2$\,GeV being the vacuum expectation value of the Higgs doublet), assuming that the ALP mass is much smaller than this scale, i.e.~$m_a\ll\mu_w$. The appropriate effective theory in this regime is called the low-energy effective field theory (LEFT), in which the heavy gauge bosons $W^\pm$ and $Z$, the Higgs boson as well as the top quark are no longer present as light degrees of freedom. 

In the presence of {\em heavy new physics\/} at a scale $\Lambda$, the Wilson coefficients of the LEFT operators receive contributions from two different sources:
\begin{itemize}
\item
matching corrections arising when the heavy SM particles are integrated out, which are suppressed by powers of $1/v^m$, and
\item
matching corrections arising when the higher-dimensional operators in the SMEFT Lagrangian are matched onto the LEFT, yielding contributions suppressed by $1/(\Lambda^n v^m)$.
\end{itemize}
At tree-level, both sets of matching conditions have been derived in \cite{Jenkins:2017jig}. Note that in the presence of new physics the LEFT provides a double expansion in $1/\Lambda$ and $1/v$, and it is by no means guaranteed that the two expansion parameters are of similar magnitude.

When the SM is extended by a {\em light new particle\/} with weak, higher-dimensional couplings to SM fields, a third source of contributions appears. In analogy to our discussion in \cite{Galda:2021hbr}, a consistent low-energy description of the SM extended by a light ALP must be based on the effective Lagrangian 
\begin{equation}
   {\cal L}_{\rm eff} = {\cal L}_{\rm LEFT} + {\cal L}_{\rm ALP} + {\cal L}_{\rm SM+ALP} \,,
\end{equation}
where both the LEFT Lagrangian and the interactions of SM fields with the ALP must be written in terms of operators containing only the light SM fields. The presence of an ALP leads to 
\begin{itemize}
\item
inhomogeneous source terms in the RG evolution equations for the LEFT operators, which generate ALP-induced contributions to these coefficients via RG evolution. 
\end{itemize}
In \cite{Galda:2021hbr} we have calculated the corresponding effects in SMEFT, which generate non-zero contributions to the SMEFT Wilson coefficients below the scale of Peccei--Quinn symmetry breaking. For the LEFT, we find in an analogous way 
\begin{equation}\label{ALPsources}
	\frac{d}{d\ln\mu}\,C_i^{\mathrm{LEFT}} - \gamma_{ji}^{\mathrm{LEFT}}\,C_j^{\mathrm{LEFT} } 
	= \frac{S_i}{(4\pi f)^2}\quad\quad (\mathrm{for}\,\,\mu<\mu_w) \,,
\end{equation}
where the ``ALP decay constant'' $f$ is related to the scale of global symmetry breaking via $\Lambda=4\pi f$. In the following we present the derivation of the source terms $S_i$ for operators up to dimension-six order in the minimal LEFT basis constructed in \cite{Jenkins:2017jig}. (The source terms for some dimension-8 operators are discussed in the appendix.) The one-loop anomalous-dimension matrix $\gamma^{\mathrm{LEFT}}$ has been calculated in \cite{Jenkins:2017dyc}.

When the RG equations \eqref{ALPsources} are integrated over a large scale interval, a source term $S_i$ generates a large single-logarithmic contribution to the corresponding LEFT coefficient $C_i$ at one-loop order. Furthermore, if the coefficient $C_i$ mixes with other coefficients $C_j$ under scale evolution at one-loop order, and if these other coefficients also have non-zero source terms $S_j$, this generates large double-logarithmic contributions to $C_i$ at two-loop order. The latter effect gives the dominant ALP contribution to $C_i$ in cases where $S_i=0$. In this way, the ALP-induced contributions to all coefficients can be worked out in a systematic way.

\section{ALP couplings below the electroweak scale} 

We consider an ALP $a$ that is a pseudoscalar gauge-singlet under the SM, whose couplings to SM fields are protected by a  classical shift symmetry. The leading interactions are mediated by dimension-five operators and have been constructed a long time ago \cite{Georgi:1986df}. In more recent work, the RG evolution equations of the ALP couplings have been derived up to two-loop order in gauge couplings \cite{Chala:2020wvs,Bauer:2020jbp}. Explicit analytic solutions to these equations were obtained, and at the electroweak scale $\mu_w$ the general effective ALP Lagrangian has been matched onto the low-energy effective ALP Lagrangian valid for $\mu<\mu_w$ \cite{Bauer:2020jbp}
\begin{equation}\label{LlowE}
\begin{aligned}
   {\cal L}_{\rm eff}^{D\le 6}
   &= \frac12 \left( \partial_\mu a\right)\!\left( \partial^\mu a\right) - \frac{m_a^2}{2}\,a^2
    + c_{GG}\,\frac{\alpha_s}{4\pi}\,\frac{a}{f}\,G_{\mu\nu}^a\,\tilde G^{\mu\nu,a}
    + c_{\gamma\gamma}\,\frac{\alpha}{4\pi}\,\frac{a}{f}\,F_{\mu\nu}\,\tilde F^{\mu\nu} \\
   &\quad\! + \frac{\partial^\mu a}{f}\spac\Big[
    \bar u_L\spac\bm{k}_U\spac\gamma_\mu\spac u_L + \bar u_R\spac\bm{k}_u\spac\gamma_\mu\spac u_R 
    + \bar d_L\spac\bm{k}_D\spac\gamma_\mu\spac d_L + \bar d_R\spac\bm{k}_d\spac\gamma_\mu\spac d_R 
    + \bar e_L\spac\bm{k}_E\spac\gamma_\mu\spac e_L + \bar e_R\spac\bm{k}_e\spac\gamma_\mu\spac e_R \Big] ,
\end{aligned}
\end{equation}
where $G_{\mu\nu}^a$ and $F_{\mu\nu}$ are the field-strength tensors of $SU(3)_c$ and $U(1)_{em}$, $\tilde G^{\mu\nu,a}$ and $\tilde F^{\mu\nu}=\frac12\,\spac\epsilon^{\mu\nu\alpha\beta}\spac F_{\alpha\beta}$ (with $\epsilon^{0123}=1$) are the corresponding dual field-strength tensors, and we define the covariant derivative as $D_\mu=\partial_\mu-i g_s\spac G_\mu^a\,t^a-i e\spac Q A_\mu$. An explicit mass term $m_a\ll 4\pi f$ softly breaks the classical shift symmetry. In this work, we assume that the ALP is light on the electroweak scale, such that the stronger condition $m_a\ll\mu_w$ holds. In the ALP effective theory above the electroweak scale, there exist dimension-six interactions connecting two ALP fields to two Higgs doublets \cite{Bauer:2020jbp}. No corresponding interactions exists in the low-energy Lagrangian \eqref{LlowE}, in which the Higgs-field has been integrated out. Higher-order effective ALP interactions start at dimension-7 order and are suppressed by $1/f^3$.

The quantities $\bm{k}_i$ are hermitian matrices in generation space, defined in the fermion mass basis. The two matrices relating to left-handed quarks satisfy the relation $\bm{k}_D = \bm{V}^\dagger \bm{k}_U\bm{V}$, where $\bm{V}$ denotes the Cabibbo--Kobayashi--Maskawa matrix. The fermion fields are 3-component vectors in generation space; however, since the top quark has been integrated out, we have $u_{L,R}=P_{L,R}\,(u~c\spac~0)^T$ in the up-type quark sector. Integrating by parts and using the well-known equation for the chiral anomaly, the Lagrangian \eqref{LlowE} can be rewritten in the equivalent form \cite{Bauer:2020jbp,Biekotter:2023mpd}
\begin{equation}\label{LlowEalt}
\begin{aligned}
   {\cal L}_{\rm eff}^{D\le 6}(\mu<\mu_w)
   &= \frac12 \left( \partial_\mu a\right)\!\left( \partial^\mu a\right) - \frac{m_a^2}{2}\,a^2
    + \tilde c_{GG}\,\frac{\alpha_s}{4\pi}\,\frac{a}{f}\,G_{\mu\nu}^a\,\tilde G^{\mu\nu,a}
    + \tilde c_{\gamma\gamma}\,\frac{\alpha}{4\pi}\,\frac{a}{f}\,F_{\mu\nu}\,\tilde F^{\mu\nu} \\
   &\quad - \frac{i\spac a}{f}\,\Big[ \bar u_L\spac\tilde{\bm{M}}_u\spac u_R 
    + \bar d_L\spac\tilde{\bm{M}}_d\,d_R + \bar e_L\spac\tilde{\bm{M}}_e\,e_R - \text{h.c.} \Big] \\
   &\quad + \frac{a^2}{2f^2}\,\Big[ \bar u_L\spac\bm{M}_u'\spac u_R 
    + \bar d_L\spac\bm{M}_d'\,d_R + \bar e_L\spac\bm{M}_e'\,e_R + \text{h.c.} \Big] \,,
\end{aligned}
\end{equation}
where
\begin{equation}\label{eq:Deltaf}
\begin{aligned}
   \tilde{\bm{M}}_u = \bm{m}_u\spac\bm{k}_u - \bm{k}_U\spac\bm{m}_u \,, \qquad
   \tilde{\bm{M}}_d = \bm{m}_d\,\bm{k}_d - \bm{k}_D\spac\bm{m}_d \,, \qquad
   \tilde{\bm{M}}_e = \bm{m}_e\,\bm{k}_e - \bm{k}_E\,\bm{m}_e
\end{aligned}
\end{equation}
and likewise
\begin{equation}
   \bm{M}_u' = \bm{m}_u\spac\bm{k}_u^2 -2\spac\bm{k}_U\spac\bm{m}_u\spac\bm{k}_u + \bm{k}_U^2\spac\bm{m}_u \quad
   \text{etc.}
\end{equation}
are matrices in generation space describing the ALP--fermion couplings, $\bm{m}_f$ are the diagonal mass matrices for the SM fermions, and
\begin{equation}\label{eq:tildecVV}
   \tilde c_{GG} = c_{GG} + \frac12 \sum_{q\ne t}\spac c_{qq} \,, \qquad
   \tilde c_{\gamma\gamma} = c_{\gamma\gamma} + \sum_{f\ne t}\spac N_c^f\spac Q_f^2\,c_{ff}
\end{equation}
are the modified ALP--boson couplings in the new basis. The sums extend over all light quark/fermion mass eigenstates, $Q^f$ are the electric charges of the fermions, and $N_c^f$ denotes their color multiplicity ($N_c^q=3$ for quarks and $N_c^l=1$ for leptons). Notice the appearance of Yukawa-like dimension-six interactions in the effective Lagrangian \eqref{LlowEalt} in the new basis. In the LEFT, a consistent power counting requires that one counts the masses of the light SM fermions as parametrically suppressed compared with the electroweak scale, $m_f\ll v$, and these fermion masses should be included in the definition of the operators. Likewise, we will treat the ALP mass parameter $m_a$ as a low-energy scale.

The parameters
\begin{equation}
   c_{u_i u_i} \equiv (k_u)_{ii} - (k_U)_{ii} \,, \qquad
   c_{d_i d_i} \equiv (k_d)_{ii} - (k_D)_{ii} \,, \qquad
   c_{e_i e_i} \equiv (k_e)_{ii} - (k_E)_{ii}
\end{equation}
with generation index $i=1,2,3$ give the leading contributions to the flavor-diagonal ALP--fermion couplings. In terms of these parameters, the interactions involving fermions in \eqref{LlowEalt} can be rewritten in the form
\begin{equation}
   {\cal L}_{\rm eff}^{D\le 6}(\mu<\mu_w)
   \ni - \frac{a}{f}\,\sum_{f\ne t}\spac m_f\,c_{ff}\,\bar f\spac i\gamma_5 f 
    + \text{flavor-changing terms} \,+\, \text{dim-6 interactions.}
\end{equation}
Since the neutrinos are massless in the SM, it was legitimate to leave out ALP couplings to neutrinos in \eqref{LlowE}. 

The ALP--boson couplings $c_{GG}$ and $c_{\gamma\gamma}$  in the original Lagrangian \eqref{LlowE} are scale invariant. The RG equations for the flavor-diagonal ALP--fermion couplings are given by \cite{Bauer:2020jbp}
\begin{equation}\label{eq:kRGE}
   \frac{d\spac c_{qq}}{d\ln\mu}
   = \frac{2\spac\alpha_s^2}{\pi^2}\,\tilde c_{GG}
    + \frac{3\spac\alpha^2}{2\pi^2}\,Q_q^2\,\tilde c_{\gamma\gamma} \,, \qquad
   \frac{d\spac c_{ll}}{d\ln\mu}
   = \frac{3\spac\alpha^2}{2\pi^2}\,\tilde c_{\gamma\gamma} \,, 
\end{equation}
where for simplicity we neglect contributions involving two powers of small Yukawa couplings of the light SM fermions, which is an excellent approximation. The flavor off-diagonal ALP--fermion couplings are scale invariant in this approximation. Here and below all renormalized couplings are defined as running couplings evaluated at the scale $\mu$. 

In the present work, like in \cite{Galda:2021hbr}, we find it convenient to use the alternative form of the effective Lagrangian shown in \eqref{LlowEalt}, because this gives rise to fewer redundant operators in the matching onto the LEFT Lagrangian. It is conventional in parts of the SMEFT and LEFT literature not to write out factors of gauge couplings in operators containing field-strength tensors, and we therefore introduce new parameters 
\begin{equation}\label{eq:ALPbosonnew}
   C_{GG} = \frac{\alpha_s}{4\pi}\,\tilde c_{GG} \,, \qquad
   C_{\gamma\gamma} = \frac{\alpha}{4\pi}\,\tilde c_{\gamma\gamma}
\end{equation}
when deriving the source terms in \eqref{ALPsources}. In terms of these new coefficients, the evolution equations for the relevant ALP couplings take the form
\begin{equation}\label{eq:kRGEnew}
   \frac{d\spac c_{qq}}{d\ln\mu}
   = \frac{8\spac\alpha_s}{\pi}\,C_{GG} + \frac{6\spac\alpha}{\pi}\,Q_q^2\,C_{\gamma\gamma} \,, \qquad
   \frac{d\spac c_{ll}}{d\ln\mu}
   = \frac{6\spac\alpha}{\pi}\,C_{\gamma\gamma} \,, 
\end{equation}
and 
\begin{equation}\label{eq:CgaugeRunning}
\begin{aligned}
   \frac{d\spac C_{GG}}{d\ln\mu} 
   &= - 2\spac\beta_{\rm QCD}\,C_{GG}
    + \frac{\alpha_s}{4\pi}\,\frac12 \sum_{q\ne t}\spac\frac{d\spac c_{qq}}{d\ln\mu} &
   &= - \beta_0^{\rm QCD}\,\frac{\alpha_s}{2\pi}\,C_{GG} + \mathcal{O}(\alpha_s^2) \,, \\ 
   \frac{d\spac C_{\gamma\gamma}}{d\ln\mu} 
   &= - 2\spac\beta_{\rm QED}\,C_{\gamma\gamma}
    + \frac{\alpha}{4\pi}\,\sum_{f\ne t}\spac N_c^f\spac Q_f^2\,\frac{d\spac c_{ff}}{d\ln\mu} \hspace{-3.5mm} &
   &= - \beta_0^{\rm QED}\,\frac{\alpha}{2\pi}\,C_{\gamma\gamma}
    + \mathcal{O}(\alpha\spac\alpha_s) \,,\end{aligned}
\end{equation}
where we have defined the QCD and QED $\beta$-functions as
\begin{equation}\label{betadef}
   \frac{d\spac\alpha_s}{d\ln\mu} = - 2\spac\alpha_s\,\beta_{\rm QCD} \,; \qquad
   \beta_{\rm QCD} = \beta_0^{\rm QCD}\,\frac{\alpha_s}{4\pi} + \dots \,,
\end{equation}
and similarly for the QED coupling. The one-loop coefficients are given by
\begin{equation}
   \beta_0^{\rm QCD} = \frac{11}{3}\spac N_c - \frac23\,n_q \,, \qquad
   \beta_0^{\rm QED} = - \frac43\,\sum_{f\ne t}\spac N_c^f\spac Q_f^2 \,,
\end{equation}
with $n_q=5$. If the factorization scale $\mu$ is lowered below the mass scale of another fermion (the next lightest one would be the bottom quark), then that fermion is integrated out from the effective theory. Consequently, $n_q$ is then the number of light quark flavors with mass below the scale $\mu$, while the sum over $f$ includes only fermions with mass below $\mu$. The same prescription must be implemented in the sums in \eqref{eq:tildecVV}.

\section{Green's functions requiring LEFT counterterms}

In order to find the ALP source terms to the RG evolution equations of the LEFT Wilson coefficients, all UV-divergent one-loop Green's functions requiring local LEFT operators as counterterms need to be computed. Choosing a Green's basis of operators \cite{Gherardi:2020det}, which is complete before applying the equations of motion to project onto the minimal LEFT basis of \cite{Jenkins:2017jig}, allows us to restrict these computations to one-particle irreducible, connected Green's functions. The relevant (redundant) operator classes are in essence those considered for the case of the SMEFT in \cite{Galda:2021hbr} but with the Higgs field removed. Operators are built out of fermion mass eigenstates, photon and gluon fields, as well as covariant derivatives. Up to dimension-six order, there are three operator classes:
\begin{itemize}
\item 
pure gauge-boson operators: $X^3$, $X^2 D^2$, $X^2$
\item 
single fermion-current operators: $\psi^2 X D$, $\psi^2 X$, $\psi^2 D^3$, $\psi^2 D^2$, $\psi^2 D$, $\psi^2$
\item 
four-fermion operators: $\psi^4$
\end{itemize}
Here $X$ denotes a (normal or dual) field-strength tensor and $\psi$ a SM fermion. For all relevant Green's functions with virtual ALP exchange, we write the sum over Feynman diagrams as
\begin{equation}
   \sum_i D_i^{\mathrm{ALP}} \equiv \frac{i\mathcal{A}}{(4\pi f)^2} \,,
\end{equation}
where the coefficient of the UV $1/\epsilon$ pole, with $\epsilon=(4-d)/2$ being the dimensional regulator, is expressed in terms of matrix elements $\langle Q_i\rangle$ of the operators in the Green's basis. In order to remove these poles in the process of renormalization, the LEFT operators are required as counterterms. This, in turn, implies that the ALP couplings appear as source terms in the RG evolution equations for the LEFT Wilson coefficients, as shown in \eqref{ALPsources}. 

\subsection{Pure gauge-boson operators}

\begin{figure}
\centering
\includegraphics[scale=1.08]{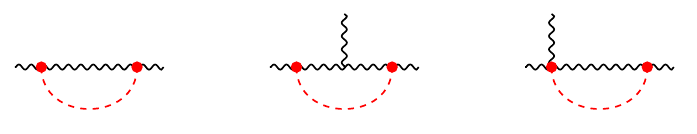}
\caption{Representative one-loop Feynman diagrams with ALP exchange, which require operators in the classes $X^3$, $X^2D^2$ and $X^2$ as counterterms. The second and third graph, which involve the three gauge-boson vertex, exist only for gluons and are absent for photons.}
\label{fig:GaugeBoson}
\end{figure}

The relevant Feynman diagrams with virtual ALP exchange requiring pure gauge-boson LEFT operators as counterterms are shown in Figure~\ref{fig:GaugeBoson}. Here and below, a red dashed line represents an ALP propagator, while red dots mark the $1/f$-suppressed ALP couplings to SM particles. Wavy lines denote gluons or photons. In order to determine the coefficients of the counter\-terms we study both the two-boson and three-boson Green’s functions with off-shell external momenta. The three-boson amplitudes only exist for non-abelian gauge fields. For the purpose of expressing the $1/\epsilon$ pole terms in these amplitudes in terms of matrix elements of LEFT operators, it is necessary to define the redundant operators
\begin{equation}\label{eq:redundantBoson}
\begin{aligned}
   \widehat{Q}_{G,2} 
   &= (D^\rho G_{\rho\mu})^a (D_\omega G^{\omega\mu})^a \,, \\
   \widehat{Q}_{\gamma,2} 
   &= (\partial^\rho F_{\rho\mu}) (\partial_\omega F^{\omega\mu}) \,,
\end{aligned}
\end{equation}
which will later be projected onto the minimal LEFT basis using the equations of motion. The divergent terms in the amplitudes can then be written in the form 
\begin{equation}\label{eq:PureGaugeBoson}
\begin{aligned}
   \mathcal{A}(gg(g)) 
   &= - \frac{C^2_{GG}}{\epsilon} \left[ 4 g_s\spac\langle Q_G\rangle 
    + \frac43\spac\langle\widehat{Q}_{G,2}\rangle - 2 m_a^2\spac\langle G_{\mu\nu}^a G^{\mu\nu,a}\rangle \right]
    +\text{finite} \,, \\
   \mathcal{A}(\gamma\gamma) 
   &= - \frac{C^2_{\gamma\gamma}}{\epsilon} \left[  \frac43\spac\langle\widehat{Q}_{\gamma,2}\rangle 
    - 2 m_a^2\spac\langle F_{\mu\nu} F^{\mu\nu}\rangle \right] + \text{finite} \,,
\end{aligned}
\end{equation}
where the Weinberg operator is defined as $Q_G=f^{abc}\,G_\mu^{\nu,a}\spac G_\nu^{\rho,b}\spac G_\rho^{\mu,c}$. These results are analogous to the ones found in \cite{Galda:2021hbr}. In both cases, the presence of the contributions proportional to the ALP mass parameter $m_a^2$ leads to a wave-function renormalization of the gauge fields, which affects the scale evolution of the running couplings $\alpha_s(\mu)$ and $\alpha(\mu)$. This will be discussed in detail in Section~\ref{sec:4}. 

The redundant operators defined in \eqref{eq:redundantBoson} can be projected onto the minimal LEFT basis by using the equations of motion 
\begin{equation}
	(D^\rho G_{\rho\mu})^a
	= g_s \sum_{q\ne t}\spac\bar q\spac\gamma_\mu\spac t^a q \,, \qquad
 	\partial^\rho F_{\rho\mu} 
	= e \sum_{f\ne t}\spac Q_f\,\bar f\spac\gamma_\mu f \,,
\end{equation}
where as before the sums extend over all light quark/fermion mass eigenstates. Transformed into LEFT notation \cite{Jenkins:2017jig}, we obtain the relations
\begin{equation}\label{EOM1}
\begin{aligned}
   \widehat{Q}_{G,2} 
   &\cong g_s^2\,\bigg[\, \frac12 \left( \big[ Q_{uu}^{V,LL} \big]_{pssp} + \big[ Q_{uu}^{V,RR} \big]_{pssp}
    + \big[ Q_{dd}^{V,LL} \big]_{pssp} + \big[ Q_{dd}^{V,RR} \big]_{pssp} \right) \\
   &\hspace{1.1cm} - \frac{1}{2N_c} \left( \big[ Q_{uu}^{V,LL} \big]_{ppss} + \big[ Q_{uu}^{V,RR} \big]_{ppss} 
    + \big[ Q_{dd}^{V,LL} \big]_{ppss} + \big[ Q_{dd}^{V,RR} \big]_{ppss} \right) \\
   &\hspace{1.1cm} + 2 \left( \big[ Q_{uu}^{V8,LR} \big]_{ppss} + \big[ Q_{dd}^{V8,LR} \big]_{ppss} \right) \\
   &\hspace{1.1cm} + 2 \left( \big[ Q_{ud}^{V8,LL} \big]_{ppss} + \big[ \big[ Q_{ud}^{V8,RR} \big]_{ppss} 
    + \big[ Q_{ud}^{V8,LR} \big] + \big[ Q_{du}^{V8,LR} \big] \right) \bigg] \,,
\end{aligned} 
\end{equation}
and
\begin{equation}\label{EOM2}
\begin{aligned}
   \widehat{Q}_{\gamma,2} 
   &\cong e^2\,\bigg[\, Q_u^2 \left( \big[ Q_{uu}^{V,LL} \big]_{ppss} + \big[ Q_{uu}^{V,RR} \big]_{ppss}
    + 2\spac\big[ Q_{uu}^{V1,LR} \big]_{ppss} \right) \\
   &\hspace{1.1cm} + Q_d^2 \left( \big[ Q_{dd}^{V,LL} \big]_{ppss} + \big[ Q_{dd}^{V,RR} \big]_{ppss}
    + 2\spac\big[ Q_{dd}^{V1,LR} \big]_{ppss} \right) \\
      &\hspace{1.1cm} + Q_e^2 \left( \big[ Q_{ee}^{V,LL} \big]_{ppss} + \big[ Q_{ee}^{V,RR} \big]_{ppss}
    + 2\spac\big[ Q_{ee}^{V,LR} \big]_{ppss} \right) \\
   &\hspace{1.1cm} + 2\spac Q_u Q_d \left( \big[ Q_{ud}^{V1,LL} \big]_{ppss} + \big[ Q_{ud}^{V1,RR} \big]_{ppss}
    + \big[ Q_{ud}^{V1,LR} \big]_{ppss} + \big[ Q_{du}^{V1,LR} \big]_{ppss} \right) \\
   &\hspace{1.1cm} + 2\spac Q_e Q_u \left( \big[ Q_{eu}^{V,LL} \big]_{ppss} + \big[ Q_{eu}^{V,RR} \big]_{ppss}
    + \big[ Q_{eu}^{V,LR} \big]_{ppss} + \big[ Q_{ue}^{V,LR} \big]_{ppss} \right) \\
   &\hspace{1.1cm} + 2\spac Q_e Q_d \left( \big[ Q_{ed}^{V,LL} \big]_{ppss} + \big[ Q_{ed}^{V,RR} \big]_{ppss}
    + \big[ Q_{ed}^{V,LR} \big]_{ppss} + \big[ Q_{de}^{V,LR} \big]_{ppss} \right) \bigg] \,,
\end{aligned} 
\end{equation}
in which a sum over the flavor indices $p,s\in\{1,2,3\}$ is implied, where the index 3 must be omitted if it refers to an up-type quark.

\subsection{Single fermion-current operators}

Operators in the classes $\psi^2 X D$, $\psi^2 D^3$, and $\psi^2 D^2$ are not generated by ALP exchange at one-loop order if one adopts the alternative form of the effective ALP Lagrangian in \eqref{LlowEalt}. The final results for the source terms are, of course, independent of that choice. The remaining operator classes can be grouped into the dipole operators $\psi^2 X$, and the operators $\psi^2 D$, $\psi^2$ contributing to the fermion two-point functions and hence to wave-function and mass renormalization effects.

\subsubsection*{Dipole operators}

\begin{figure}
\centering
\includegraphics[scale=1.13]{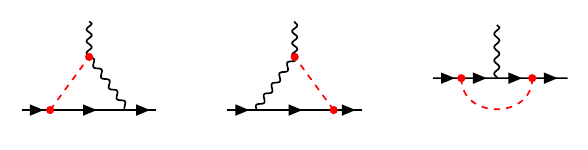}
\caption{One-loop Feynman diagrams with ALP exchange, which require dipole operators in the class $(\bar L R)X$ as counterterms. The third diagram does not give rise to a divergent contribution.}
\label{fig:Dipole}
\end{figure}

The dipole operators are chirality-changing, and for concreteness we focus on operators in the class $(\bar L R)X$ in the notation of \cite{Jenkins:2017jig}. The relevant Feynman diagrams are shown in Figure~\ref{fig:Dipole}, where the fermion can be a quark or a charged lepton. The diagrams exist for both external gluons and photons. They correspond to the graphs evaluated for the SMEFT case in \cite{Galda:2021hbr}, but with the external Higgs field replaced by its vacuum expectation value. Note that the third diagram yields contributions to the dipole amplitudes that are UV finite. We find that the UV poles in the amplitudes obtained from these graphs read (with $q=u,d$)
\begin{equation}\label{dipoleq}
   \mathcal{A}(\bar q_{L,p}\spac q_{R,r}\spac g) 
   = - \frac{2 g_s}{\epsilon}\,\big( \tilde{\bm{M}}_q \big)_{pr}\,C_{GG}\,\big[ \langle Q_{qG} \rangle \big]_{pr}
    + \text{finite} \,, 
\end{equation}
and (with $f=u,d,e$)
\begin{equation}\label{dipolef}
   \mathcal{A}(\bar f_{L,p}\spac f_{R,r}\spac\gamma) 
   = - \frac{2\spac Q_f\spac e}{\epsilon}\,\big( \tilde{\bm{M}}_f \big)_{pr}\,C_{\gamma\gamma}\,
    \big[ \langle Q_{f\gamma} \rangle \big]_{pr} + \text{finite} \,, 
\end{equation}
where $p,r$ are generation indices, and no summation over these indices is implied. The flavor matrices $\tilde{\bm{M}}_f$ have been defined in \eqref{eq:Deltaf}.

\subsubsection*{Fermion two-point functions}

ALP exchange also yields new contributions to the fermion two-point functions, arising from the diagrams
\begin{figure}[h]
\centering
\includegraphics[scale=1.13]{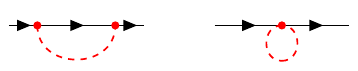}
\end{figure}

\noindent
where the tadpole graph contains an insertion of the dimension-six operators shown in the last line of \eqref{LlowEalt}. The resulting UV-divergent contributions to the fermion self-energies (written as a matrix in generation space) can be decomposed as
\begin{equation}
   \bm{\Sigma}^{(f)}(p)
   = \bm{\Sigma}_{LL}^{(f)} + \bm{\Sigma}_{RR}^{(f)} + \bm{\Sigma}_{LR}^{(f)} + \bm{\Sigma}_{RL}^{(f)} \,,
\end{equation}
where (with $f=u,d,e$)
\begin{equation}\label{selfenergies}
\begin{aligned}
   \bm{\Sigma}_{LL}^{(f)}
   &= \frac{1}{(4\pi f)^2}\,\frac{1}{\epsilon}
    \left( - \frac{\rlap{\spac/}{p}}{2} \right) P_L\,\tilde{\bm{M}}_f\spac\tilde{\bm{M}}_f^\dagger 
    + \text{finite} \,, \\
   \bm{\Sigma}_{RR}^{(f)}
   &= \frac{1}{(4\pi f)^2}\,\frac{1}{\epsilon}
    \left( - \frac{\rlap{\spac/}{p}}{2} \right) P_R\,\tilde{\bm{M}}_f^\dagger\spac\tilde{\bm{M}}_f 
    + \text{finite} \,, \\
   \bm{\Sigma}_{LR}^{(f)}
   &= \frac{1}{(4\pi f)^2}\,\frac{1}{\epsilon}\,P_R\,\Big( 
    \tilde{\bm{M}}_f\,\bm{m}_f\spac\tilde{\bm{M}}_f 
    + \frac{m_a^2}{2}\,\bm{M}_f' \Big) + \text{finite} \,, \\
   \bm{\Sigma}_{RL}^{(f)}
   &= \frac{1}{(4\pi f)^2}\,\frac{1}{\epsilon}\,P_L\,\Big( 
    \tilde{\bm{M}}_f^\dagger\,\bm{m}_f\spac\tilde{\bm{M}}_f^\dagger 
    + \frac{m_a^2}{2}\,\bm{M}_f^{\prime\,\dagger} \Big) + \text{finite} \,.
\end{aligned}
\end{equation}
These contributions must be combined with the SM expression for the self-energy and then brought to canonical (and diagonalized) form using field redefinitions. 

\subsection{Four-fermion operators}
\label{subsec:3.3}

\begin{figure}
\centering
\includegraphics[scale=1.13]{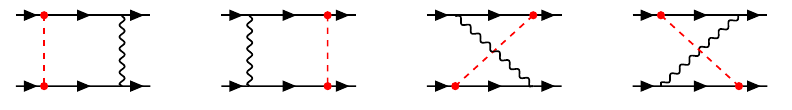}
\caption{One-loop Feynman diagrams with ALP exchange contributing to the four-fermion amplitudes. All graphs are UV finite.}
\label{fig:FourFermion}
\end{figure}

The Feynman diagrams which could in principle require local dimension-six LEFT operators in the classes $(\bar{L}L)(\bar{L}L)$, $(\bar{R}R)(\bar{R}R)$, and $(\bar{L}L)(\bar{R}R)$ as counterterms are depicted in Figure~\ref{fig:FourFermion}. However, it can be readily seen that these diagrams are free of UV divergences, because we work with the alternative form of the effective ALP Lagrangian in \eqref{LlowEalt}, featuring non-derivative ALP--fermion couplings. Nevertheless, non-vanishing source terms for the four-fermion operators arise when we apply the equations of motion to eliminate the redundant operators in \eqref{EOM1} and \eqref{EOM2}. These source terms are proportional to $C_{GG}^2$ or $C_{\gamma\gamma}^2$, respectively.

In the corresponding analysis of four-fermion operators in the SMEFT, the source terms of the four-fermion operators also contain contributions quadratic in the ALP--fermion couplings \cite{Galda:2021hbr}. Such terms also exist in the LEFT, but they contribute at higher orders in power counting. The reason is that a consistent LEFT power counting requires that one treats the SM Yukawa matrices as power-suppressed quantities of $\mathcal{O}(m_f/v)$ \cite{Jenkins:2017jig}, where $m_f\ll v$ are the masses of the light SM fermions. Consider now a diagram of the type
\begin{figure}[h]
\centering
\includegraphics[scale=1.13]{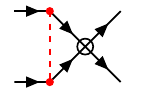}
\end{figure}

\noindent
in which the symbol $\otimes$ denotes an effective four-fermion vertex in the LEFT, which arises when the heavy electroweak gauge bosons $W^\pm$, $Z$ and the Higgs boson are integrated out. Such vertices appear at dimension-six ($\sim\!1/v^2$) and dimension-8 order ($\sim\!m_f\spac m_{f'}/v^4$), respectively. The ALP exchange leads to an additional suppression factor $1/\Lambda^2=1/(4\pi f)^2$. Hence, the contributions arising from $W^\pm $ or $Z$ exchange appear first at dimension-8 order in the LEFT and scale like $1/(\Lambda^2\spac v^2)$, while the Higgs-induced contributions appear first at dimension-10 order and scale like $1/(\Lambda^2\spac v^4)$. As an explicit example, we discuss in the appendix the dimension-8 source terms of the four-fermion operators with flavor structure $(\bar u\spac u)(\bar e\spac e)$ originating from $Z$-boson exchange.

\section{Derivation of the source terms}
\label{sec:4}

The ALP-induced UV-divergent one-loop Green's functions presented above require local LEFT operators as counterterms. Concretely, if a Green's function contains a $1/\epsilon$ pole term of the form
\begin{equation}
	\frac{\xi_i}{(4\pi f)^2}\,\frac{1}{\epsilon}\,\langle Q_i \rangle 
\end{equation}
with some LEFT operator $Q_i$, then the bare Wilson coefficient of this operator must contain a counterterm contribution absorbing this divergence, i.e., it must be of the form
\begin{equation}
	C_{i,0} \ni - \frac{\xi_i}{(4\pi f)^2} \left( \frac{1}{\epsilon} + \ln\frac{\mu^2}{M_{\rm UV}^2} + \dots  \right) ,
\end{equation}
where $M_{\rm UV}$ is some UV scale. The RG evolution equation of the correspnding {\em renormalized\/} Wilson coefficient thus receives an inhomogeneous contribution of the form
\begin{equation}
	\frac{d}{d\ln\mu}\,C_i(\mu) \ni - \frac{2\spac\xi_i}{(4\pi f)^2} \,,
\end{equation}
where the coefficient $\xi_i$ is proportional to some ALP couplings. In the notation of \eqref{ALPsources}, this corresponds to the source term $S_i=-2\spac\xi_i$. In this way, the Wilson coefficients of LEFT operators are generated or modified in the presence of an ALP, irrespective of the presence of additional new heavy degrees of freedom beyond the SM. The following list shows the ALP source terms generated from one-loop ALP exchange.

\subsubsection*{Pure gauge-boson operators}

From \eqref{eq:PureGaugeBoson}, we find that the source terms for the dimension-six LEFT operators in class $X^3$ are
\begin{equation}
	S_G = 8 g_s\spac C_{GG}^2 \,, \qquad S_{\widetilde G} = 0 \,,
\end{equation}
which is the same result we found in the SMEFT \cite{Galda:2021hbr}. 

\subsubsection*{Dipole operators}

The source terms for the dipole operators in class $(\bar LR)X$ are derived from \eqref{dipoleq} and \eqref{dipolef}. Using a compact matrix notation, we obtain for the chromomagnetic dipole operators
\begin{equation}
\begin{aligned}
   \bm{S}_{uG}
   &= 4\spac g_s\spac C_{GG}\,\tilde{\bm{M}}_u \,, \\
   \bm{S}_{dG}
   &= 4\spac g_s\spac C_{GG}\,\tilde{\bm{M}}_d \,,
\end{aligned}
\end{equation}
and for the electromagnetic ones
\begin{equation}\label{eq:Segamma}
\begin{aligned}
   \bm{S}_{u\gamma}
   &= \frac83\,e\,C_{\gamma\gamma} \,\tilde{\bm{M}}_u \,, \\
   \bm{S}_{d\gamma}
   &= - \frac43\,e\,C_{\gamma\gamma}\,\tilde{\bm{M}}_d \,, \\[2mm]
   \bm{S}_{e\gamma}
   &= - 4\spac e\,C_{\gamma\gamma}\,\tilde{\bm{M}}_e \,.
\end{aligned}
\end{equation}
For models in which the ALP--fermion couplings are flavor diagonal in the mass basis, the matrices $\tilde{\bm{M}}_f$ are diagonal, with entries given by $m_f\spac c_{ff}$ for each fermion mass eigenstate.

\subsubsection*{Four-fermion operators}

At one-loop order, all source terms for four-fermion operators descent from the elimination of the redundant operators in \eqref{EOM1} and \eqref{EOM2}. We obtain non-vanishing source terms in the following three operator classes.

\vspace{3mm}
\noindent
\underline{Classes $(\bar LL)(\bar LL)$ and $(\bar RR)(\bar RR)$:}\\[2mm]
The same-chirality LEFT operators receive the source terms
\begin{equation}
\begin{aligned}
   \big[ S_{ee}^{V,LL} \big]_{prst} &= \big[ S_{ee}^{V,RR} \big]_{prst} &\hspace{-3.5mm}
   &= \frac83\,e^2\spac C_{\gamma\gamma}^2\,\delta_{pr}\,\delta_{st} \,, \\
   \big [S_{eu}^{V,LL} \big]_{prst} &= \big [S_{eu}^{V,RR} \big]_{prst} &\hspace{-3.5mm}
   &=  - \frac{32}{9}\,e^2\spac C_{\gamma\gamma}^2\,\delta_{pr}\,\delta_{st} \,, \\
   \big[ S_{ed}^{V,LL} \big]_{prst} &= \big[ S_{ed}^{V,RR} \big]_{prst} &\hspace{-3.5mm}
   &= \frac{16}{9}\,e^2\spac C_{\gamma\gamma}^2\,\delta_{pr}\,\delta_{st} \,, \\
   \big[ S_{uu}^{V,LL} \big]_{prst} &= \big[ S_{uu}^{V,RR} \big]_{prst} &\hspace{-3.5mm}
   &= \frac{32}{27}\,e^2\spac C_{\gamma\gamma}^2\,\delta_{pr}\,\delta_{st}
    + \frac43\,g_s^2\spac C_{GG}^2 \left( \delta_{pt}\,\delta_{rs} 
    - \frac{1}{N_c}\,\delta_{pr}\,\delta_{st} \right) , \\
   \big[ S_{dd}^{V,LL} \big]_{prst} &= \big[ S_{dd}^{V,RR} \big]_{prst} &\hspace{-3.5mm}
   &=  \frac{8}{27}\,e^2\spac C_{\gamma\gamma}^2\,\delta_{pr}\,\delta_{st}
	+ \frac43\,g_s^2\spac C_{GG}^2 \left( \delta_{pt}\,\delta_{rs} 
    - \frac{1}{N_c}\,\delta_{pr}\,\delta_{st} \right) , \\
   \big[ S_{ud}^{V1,LL} \big]_{prst} &= \big[ S_{ud}^{V1,RR} \big]_{prst} &\hspace{-3.5mm}
   &= - \frac{32}{27}\,e^2\spac C_{\gamma\gamma}^2\,\delta_{pr}\,\delta_{st} \,, \\
   \big[ S_{ud}^{V8,LL} \big]_{prst} &= \big[ S_{ud}^{V8,RR} \big]_{prst} &\hspace{-3.5mm}
   &= \frac{16}{3}\,g_s^2\spac C_{GG}^2\,\delta_{pr}\,\delta_{st} \,.
\end{aligned}
\end{equation}

\vspace{3mm}
\noindent
\underline{Class $(\bar LL)(\bar RR)$:}\\[2mm]
The mixed-chirality LEFT operators receive the source terms
\begin{align}\label{eq:Smixed}
   \big[ S_{ee}^{V,LR} \big]_{prst} 
   &= \frac{16}{3}\,e^2\spac C_{\gamma\gamma}^2\,\delta_{pr}\,\delta_{st} \,, \notag\\
   \big[ S_{eu}^{V,LR} \big]_{prst} 
   &= - \frac{32}{9}\,e^2\spac C_{\gamma\gamma}^2\,\delta_{pr}\,\delta_{st} \,, \notag\\
   \big[ S_{ed}^{V,LR} \big]_{prst} 
   &= \frac{16}{9}\,e^2\spac C_{\gamma\gamma}^2\,\delta_{pr}\,\delta_{st} \,, \notag\\
   \big[ S_{ue}^{V,LR} \big]_{prst} 
   &= - \frac{32}{9}\,e^2\spac C_{\gamma\gamma}^2\,\delta_{pr}\,\delta_{st} \,, \notag\\
   \big[ S_{de}^{V,LR} \big]_{prst} 
   &= \frac{16}{9}\,e^2\spac C_{\gamma\gamma}^2\,\delta_{pr}\,\delta_{st} \,, \notag\\
   \big[ S_{uu}^{V1,LR} \big]_{prst} 
   &= \frac{64}{27}\,e^2\spac C_{\gamma\gamma}^2\,\delta_{pr}\,\delta_{st} \,, \notag\\
   \big[ S_{uu}^{V8,LR} \big]_{prst} 
   &= \frac{16}{3}\,g_s^2\spac C_{GG}^2\,\delta_{pr}\,\delta_{st} \,, \\
   \big[ S_{ud}^{V1,LR} \big]_{prst} 
   &= - \frac{32}{27}\,e^2\spac C_{\gamma\gamma}^2\,\delta_{pr}\,\delta_{st} \,, \notag\\
   \big[ S_{ud}^{V8,LR} \big]_{prst} 
   &= \frac{16}{3}\,g_s^2\spac C_{GG}^2\,\delta_{pr}\,\delta_{st} \,, \notag\\
   \big[ S_{du}^{V1,LR} \big]_{prst} 
   &= - \frac{32}{27}\,e^2\spac C_{\gamma\gamma}^2\,\delta_{pr}\,\delta_{st} \,, \notag\\
   \big[ S_{du}^{V8,LR} \big]_{prst}
    &= \frac{16}{3}\,g_s^2\spac C_{GG}^2\,\delta_{pr}\,\delta_{st} \,, \notag\\
   \big[ S_{dd}^{V1,LR} \big]_{prst} 
   &= \frac{16}{27}\,e^2\spac C_{\gamma\gamma}^2\,\delta_{pr}\,\delta_{st} \,, \notag\\
   \big[ S_{dd}^{V8,LR} \big]_{prst} 
   &= \frac{16}{3}\,g_s^2\spac C_{GG}^2\,\delta_{pr}\,\delta_{st} \,. \notag
\end{align}
LEFT operators in the classes $(\bar LR)(\bar LR)$ and $(\bar LR)(\bar RL)$ are not sourced by ALP exchange at one-loop order.

\subsubsection*{Wave-function, coupling and mass renormalization}

One-loop diagrams with a virtual ALP exchange can also generate divergent contributions to Green's functions which require the dimension-4 operators of the SM as counterterms. We have seen two occurrences of this phenomenon: the contributions proportional to $m_a^2$ in \eqref{eq:PureGaugeBoson}, and the contributions \eqref{selfenergies} to the fermion self-energies, which are proportional to powers of the light fermion masses. 

The ALP contributions to the wave-function renormalization constants of the gauge fields derived from \eqref{eq:PureGaugeBoson} read
\begin{equation}
   \delta Z_G = \frac{8\spac m_a^2}{(4\pi f)^2}\,\frac{C_{GG}^2}{\epsilon} \,, \qquad
   \delta Z_\gamma = \frac{8\spac m_a^2}{(4\pi f)^2}\,\frac{C_{\gamma\gamma}^2}{\epsilon} \,.
\end{equation}
The corresponding contributions to the $\beta$-functions of QCD and QED, as defined in \eqref{betadef}, are~\cite{Galda:2021hbr}
\begin{equation}\label{eq:ALPbeta}
\begin{aligned}
   \beta_{\rm QCD} 
   &= \beta_{\rm QCD}^{\rm SM} + \frac{8\spac m_a^2}{(4\pi f)^2}\,C_{GG}^2 \,, \\
   \beta_{\rm QED}
   &= \beta_{\rm QED}^{\rm SM} + \frac{8\spac m_a^2}{(4\pi f)^2}\,C_{\gamma\gamma}^2 \,.
\end{aligned}
\end{equation}

We will discuss the renormalization of the fermion two-point functions assuming for simplicity that the ALP--fermion couplings are flavor-diagonal in the mass basis. Then the ALP contributions to the wave-function and mass renormalization of fermion $f$ are given by 
\begin{equation}
   \delta Z_f = - \frac{m_f^2}{(4\pi f)^2}\,\frac{c_{ff}^2}{2\epsilon} \,, \qquad
   \delta m_f = \frac{m_f}{(4\pi f)^2}\,\frac{c_{ff}^2}{\epsilon} \left( m_f^2 + \frac{m_a^2}{2} \right) .
\end{equation}
The divergent contribution to the fermion mass affects the running of the fermion masses, governed by the evolution equation $dm_f/d\ln\mu=\gamma_{m_f}\,m_f$, and we find that the ALP contribution to the anomalous dimension of the running fermion mass is given by 
\begin{equation}
   \gamma_{m_f}
   = \gamma_{m_f}^{\rm SM} + \frac{2\spac m_f^2 + m_a^2}{(4\pi f)^2}\,c_{ff}^2 \,.
\end{equation}

\section{Anomalous magnetic moment of the muon}

\begin{figure}
\centering
\includegraphics[width=\textwidth]{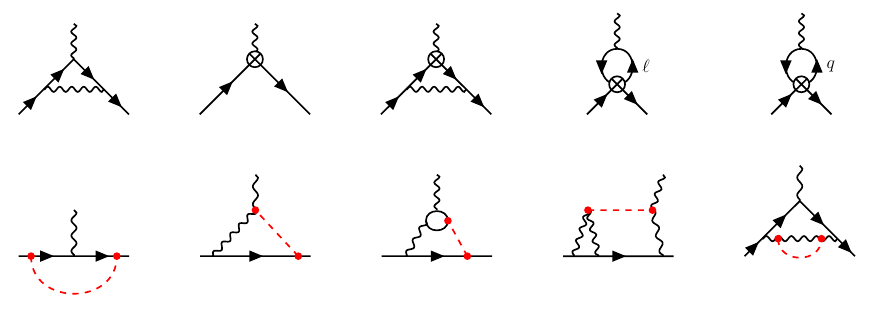}
\caption{Representative Feynman diagrams contributing to the anomalous magnetic moment of the muon at one- and partial two-loop order in the LEFT\,+\,ALP effective theory. LEFT operators are indicated by the $\otimes$ symbol (see the main text for more details).}
\label{fig:amudiags}
\end{figure}

As an important application of our formalism, we derive the ALP-induced contributions to the anomalous magnetic moment of the muon, $a_\mu\equiv\frac12\spac(g-2)_\mu$. In the LEFT, this important precision observable has been discussed in detail in \cite{Aebischer:2021uvt}. At a scale $\mu_0\sim m_\mu$, one obtains at one-loop order\footnote{In this reference a different definition of the covariant derivative is used, so that we must change the sign of the gauge couplings when transforming their expressions to ours.}
\begin{align}\label{eq:amumaster}
   a_\mu 
   &= \frac{\alpha(\mu_0)}{2\pi} - \frac{4 m_\mu}{e}\,{\Re e}\spac\big[ C_{e\gamma}(\mu_0) \big]_{22}
    \left[ 1 - \frac{\alpha}{4\pi} \left( 5\ln\frac{\mu_0^2}{m_\mu^2} + 2 \right) \right] 
    + a_\mu^{4\ell}(\mu_0) + a_\mu^{2\ell\spac 2q}(\mu_0) + a_\mu^{\rm ALP}(\mu_0) \notag\\
   &\equiv [a_\mu]^{\rm SM} + [a_\mu]^{\rm ALP} \,,
\end{align}
where the individual terms correspond to the Feynman diagrams shown in Figure~\ref{fig:amudiags}. The first contribution is the celebrated Schwinger term, while the remaining terms contain the contributions from weak-scale heavy particles and potential physics beyond the SM.\footnote{The SM prediction for $a_\mu$ is known to much higher accuracy than the one-loop result included here, see \cite{Aoyama:2020ynm} for a review. It includes the pure QED contribution up to four-loop order analytically \cite{Laporta:2017okg} and up to five-loop order in numerical form \cite{Aoyama:2012wk,Aoyama:2014sxa}, as well as detailed estimates of hadronic effects, which start at two-loop order.} 
The second term is the one-loop matrix element of the electromagnetic dipole operator in the LEFT (second and third diagrams in the first row of Figure~\ref{fig:amudiags}). The contribution
\begin{equation}\label{eq:amu4l}
   a_\mu^{4\ell}(\mu_0) 
   = m_\mu \sum_{\ell=e,\mu,\tau} \frac{m_\ell}{4\pi^2} \left[ 
    \ln\frac{\mu_0^2}{m_\ell^2}\,\,{\Re e}\spac\big[ C_{ee}^{S,RR}(\mu_0) \big]_{2\ell\ell2}
    - {\Re e}\spac\big[ C_{ee}^{V,LR}(\mu_0) \big]_{2\ell\ell2} \right]
\end{equation}
accounts for penguin diagrams involving four-lepton operators in the LEFT (fourth diagram in the first row), which arise from the exchange of $W^\pm$ and $Z$ bosons. Note that the Wilson coefficients $C_{e\gamma}$ and $C_{ee}$ contain the electroweak contributions to $a_\mu$ in the SM in addition to possible new-physics contributions, i.e.\
\begin{equation}
\begin{aligned}
   \big[ C_{e\gamma}(\mu_0) \big]_{22}^{\rm SM}
   &= \frac{G_F\spac e\spac m_\mu}{48\sqrt2\spac\pi^2} \left( -3 - 8\spac s_w^2 + 16\spac s_w^4 \right) , \\
   \big[ C_{ee}^{V,LR}(\mu_0) \big]_{2\ell\ell2}^{\rm SM}
   &= \frac{4\spac G_F}{\sqrt2} \left( 1 - 2\spac s_w^2 \right) s_w^2\,\delta_{2\ell} \,.
\end{aligned}
\end{equation}
Similarly, the contribution $a_\mu^{2\ell\spac 2q}$ arises from penguin diagrams involving a quark loop (fifth diagram in the first row). It is given in (3.4) and (3.5) of \cite{Aebischer:2021uvt} and involves the tensor couplings $C_{eu}^{T,RR}$ and $C_{ed}^{T,RR}$, which are not sourced by ALP exchange. Hence they will not play a role for our discussion. The same is true for the scalar coupling $C_{ee}^{S,RR}$ in \eqref{eq:amu4l}. 

\subsection{Low-energy ALP contribution}

In the present work our focus is on the ALP contributions to $a_\mu$, denoted by $[a_\mu]^{\rm ALP}$ in \eqref{eq:amumaster}. In addition to ALP-induced terms in the LEFT contributions, in the ALP\,+\,LEFT effective theory considered here there is an additional low-energy contribution $a_\mu^{\rm ALP}$, which arises from the diagrams shown in the second row of Figure~\ref{fig:amudiags}. It is given by \cite{Bauer:2017ris,Bauer:2021mvw}
\begin{equation}\label{eq:amuALP}
\begin{aligned}
   a_\mu^{\rm ALP}(\mu_0)
   &= \frac{m_\mu^2}{(4\pi f)^2}\,\Bigg\{ - c_{\mu\mu}^2(\mu_0)\,h_1(x_\mu)
    - 8\spac c_{\mu\mu}(\mu_0)\,C_{\gamma\gamma}(\mu_0) \left[ \ln\frac{\mu_0^2}{m_\mu^2}
    + \delta_2 + 3 - h_2(x_\mu) \right] \\
   &\hspace{2.22cm} - \frac{2\spac\alpha(\mu_0)}{\pi}\,c_{\mu\mu}(\mu_0) 
    \sum_{f\ne t}\spac N_c^f\spac Q_f^2\,c_{ff}(\mu_0) \int_0^1\!dz\,F\big(z(1-z)\spac x_f,x_\mu\big) \\[-1mm]
   &\hspace{2.22cm} + \frac{\alpha(\mu_0)}{\pi}\,C_{\gamma\gamma}^2(\mu_0)\,\Delta_{\rm LbL}(\mu_0,m_a,m_\mu) \\
   &\hspace{2.22cm} + \frac{m_\tau}{m_\mu}\,{\Re e}\spac\big( [k_e]_{23}\spac[k_E]_{23}^* \big)\,h_3(x_\tau)
    - \frac12\spac\big( |[k_e]_{12}|^2 + |[k_E]_{12}|^2 \big)\,h_4(x_\mu) \Bigg\} \,,
\end{aligned}
\end{equation}
where $x_f\equiv m_a^2/m_f^2+i0$, and the loop functions $h_i(x)$ are given by 
\begin{equation}
\begin{aligned}
   h_1(x) &= 1 + 2x + x (1-x) \ln x - 2x(3-x)\,\sqrt{\frac{x}{4-x}}\,\arccos\frac{\sqrt{x}}{2} \,, \\[-1mm]
   h_2(x) &= 1 - \frac{x}{3} + \frac{x^2}{6} \ln x + \frac{2+x}{3}\,\sqrt{x\spac(4-x)}\,\arccos\frac{\sqrt{x}}{2} \,, \\
   h_3(x) &= \frac{1-3x}{(1-x)^2} - \frac{2x^2}{(1-x)^3} \ln x \,, \\
   h_4(x) &= 1 + 2x - 2 x^2 \ln\frac{x}{x-1} \,.
\end{aligned}
\end{equation}    
They satisfy $h_i(0)=1$ as well as $h_1(x)\approx\frac{2}{x}\spac(\ln x-\frac{11}{6})$, $h_2(x)\approx(\ln x+\frac32)$, $h_3(x)\approx\frac{2}{x}\spac(\ln x-\frac32)$ and $h_4(x)\approx-\frac{2}{3x}$ for $x\gg 1$.\footnote{The expression for $h_4(x)$ develops a non-zero imaginary part for $x<1$, which reflects the fact that the electron--ALP pair in the loop can go on-shell if $(m_a+m_e)<m_\mu$. In the approximation where the electron mass is neglected, as we have done here, the result exhibits a logarithmic singularity at the threshold $m_a=m_\mu$, so the above expression should only be used as long as $m_a\ne m_\mu$.}
In the limit $m_a\ll m_\mu$ one can replace $h_i(x_\mu)\approx 1$, in which case the only remaining mass dependence of the expression inside the rectangular brackets in the first line of \eqref{eq:amuALP} is in the logarithm $\ln(\mu_0^2/m_\mu^2)$. Choosing $\mu_0\sim m_\mu$ ensures that this logarithm is of $\mathcal{O}(1)$. In the opposite limit $m_a\gg m_\mu$ one can set $h_i(x_\mu)\approx 0$ for $i=1,3,4$, whereas the expression inside the rectangular brackets simplifies to $[\ln(\mu_0^2/m_a^2)+\delta_2+\frac32]$. In this case choosing $\mu_0\sim m_a$ ensures that the relevant logarithm is of $\mathcal{O}(1)$. In general, for $m_a\gg m_\mu$ the ALP should be integrated out at a scale $\mu\sim m_a$, and below that scale the effective theory is given by the LEFT without the ALP. In our discussion below we will not specify a particular hierarchy between the ALP and muon masses, but it should be understood that in case these masses are widely separated the proper scale choice is $\mu_0\sim\max\spac(m_\mu,m_a)$. 

The scheme-dependent constant $\delta_2$ in the first line of \eqref{eq:amuALP} is related to the treatment of the Levi--Civita symbol in dimensional regularization. One finds $\delta_2=-3$ in a scheme where $\epsilon^{\mu\nu\alpha\beta}$ is treated as a $d$-dimensional object (our default choice), and $\delta_2=0$ in a scheme where it is treated as a four-dimensional quantity \cite{Bauer:2017ris}.

At two-loop order, the ALP-induced low-energy contribution to $a_\mu$ receives corrections of order $c_{\mu\mu}(\mu_0)\spac c_{ff}(\mu_0)$, for all fermions $f\ne t$, and $C_{\gamma\gamma}^2(\mu_0)$.\footnote{There are also loop corrections of order $\alpha\spac c_{\mu\mu}^2$ and $\alpha\,c_{\mu\mu}\spac C_{\gamma\gamma}$ to the one-loop terms in \eqref{eq:amuALP}, which we neglect.} 
The first effect is due to Barr--Zee diagrams \cite{Barr:1990vd} (third diagram in the second row of Figure~\ref{fig:amudiags}) and has been calculated in \cite{Buen-Abad:2021fwq,Bauer:2021mvw}. The relevant loop function is
\begin{equation}
   F(y,x) = \frac{1}{1-y} \left[ h_2(x) - h_2\bigg(\frac{x}{y}\bigg) \right] .
\end{equation}
Note that only the sum of the second and third term inside the curly brackets is independent of the choice of the effective ALP Lagrangian, i.e., it is the same irrespective of whether the calculation is performed based on the Lagrangian in \eqref{LlowE} or \eqref{LlowEalt}. The contribution proportional to $C_{\gamma\gamma}^2$ in the third line of \eqref{eq:amuALP} arises from light-by-light scattering and vacuum-polarization diagrams via ALP exchange (last two diagrams in the second row of Figure~\ref{fig:amudiags}) and is denoted by $\Delta_{\rm LbL}$ in \eqref{eq:amuALP}, a quantity which has not yet been calculated. It was analyzed in the leading logarithmic approximation (assuming a scale choice $\mu\gg m_{\mu,a}$) in \cite{Marciano:2016yhf}. Below, we will reproduce and extend the result obtained by these authors by solving the RG evolution equations in our effective theory. Therefore, our result captures the leading contributions proportional to $c_{\mu\mu}^2$, $c_{\mu\mu}\spac C_{\gamma\gamma}$ and $c_{\mu\mu}\spac c_{ff}$, while the leading contributions proportional to $C_{\gamma\gamma}^2$ are contained in the leading logarithmic approximation only. 

The two flavor off-diagonal contributions shown in the last line of \eqref{eq:amuALP} arise from the first diagram in the second row of Figure~\ref{fig:amudiags}, if the fermion in the loop is a $\tau$-lepton or an electron. In these expressions we have neglected small corrections proportional to $m_\mu/m_\tau$ or $m_e/m_\mu$, respectively. Since the flavor off-diagonal ALP--fermion couplings are scale independent, we have omitted the argument $\mu_0$ in these terms.

\subsection{RG evolution to the electroweak scale}

Our next task is to express the ALP coupling parameters and the LEFT Wilson coefficients at the low scale $\mu_0$ in terms of coupling parameters defined at the electroweak scale. To this end, we need to integrate the RG evolution equations for the ALP couplings \cite{Chala:2020wvs,Bauer:2020jbp} along with the evolution equations \eqref{ALPsources} for the LEFT Wilson coefficients from the low scale $\mu_0$ up to the scale $\mu_w$. The anomalous dimensions for the LEFT Wilson coefficients have been calculated in \cite{Jenkins:2017dyc}, while the ALP source terms have been derived in the present work. Integrating these equations numerically, one can resum large logarithmic corrections to all orders of perturbation theory. 

For the present work, we restrict ourselves to an approximate solution of the RG equations, working consistently at lowest non-trivial order in the ALP couplings. Integrating the second equation in \eqref{eq:kRGEnew} and the second equation in \eqref{eq:CgaugeRunning} we obtain 
\begin{equation}
\begin{aligned}
   c_{\mu\mu}(\mu_0)
   &\approx c_{\mu\mu}(\mu_w) - \frac{3\spac\alpha}{\pi}\,C_{\gamma\gamma}(\mu_w)\,\ln\frac{\mu_w^2}{\mu_0^2} \,, \\
   C_{\gamma\gamma}(\mu_0)
   &\approx \left[ 1 + \beta_0^{\rm QED}\,\frac{\alpha(\mu_w)}{4\pi}\,\ln\frac{\mu_w^2}{\mu_0^2} \right] 
    C_{\gamma\gamma}(\mu_w) \,, 
\end{aligned}
\end{equation}
where the correction to $C_{\gamma\gamma}$ can be neglected in our approximation because it is of higher order in $\alpha$. There is also an ALP-induced correction to the running of the QED coupling, for which we obtain from \eqref{eq:ALPbeta} 
\begin{equation}\label{eq:45}
   \big[ \alpha(\mu_0) \big]^{\rm ALP}
   \approx \alpha(\mu_w) \left[ 1 + \frac{8\spac m_a^2}{(4\pi f)^2}\,C_{\gamma\gamma}^2(\mu_w)\,
    \ln\frac{\mu_w^2}{\mu_0^2} \right] .
\end{equation}
When this expression is used in the Schwinger term in \eqref{eq:amumaster}, an ALP-induced contribution proportional to $\alpha\,C_{\gamma\gamma}^2$ is induced.

In the next step, we need to integrate the evolution equations for the relevant LEFT couplings $[C_{e\gamma}]_{22}$ and $[C_{ee}^{V,LR}]_{2\ell\ell2}$, for which the ALP source terms have been given in \eqref{eq:Segamma} and \eqref{eq:Smixed}, respectively. The dipole coefficient $C_{e\gamma}$ mixes under RG evolution with itself, with the coefficients $C_{eu}^{T,RR}$, $C_{ed}^{T,RR}$, $C_{ee}^{S,RR}$, for which there are no ALP source terms, and with squares of the dipole coefficients $C_{e\gamma}$, $C_{u\gamma}$ and $C_{d\gamma}$ times the mass of a light SM fermion \cite{Jenkins:2017dyc}. To the order we are working, these mixing effects do not give rise to relevant contributions. The four-fermion coefficient $C_{ee}^{V,LR}$ enters the expression for $a_\mu^{4\ell}$ in \eqref{eq:amu4l} at one-loop order, and hence we only need the ALP contribution obtained from integrating the source term $S_{ee}^{V,LR}$ in \eqref{eq:Smixed}. We obtain
\begin{equation}
\begin{aligned}
   \big[ C_{e\gamma}(\mu_0) \big]_{22}^{\rm ALP}
   &\approx \big[ C_{e\gamma}(\mu_w) \big]_{22}^{\rm ALP}
    + \frac{2e\spac m_\mu}{(4\pi f)^2}\,C_{\gamma\gamma}(\mu_w) \left[ c_{\mu\mu}(\mu_w)\,\ln\frac{\mu_w^2}{\mu_0^2} 
    - \frac{3\spac\alpha}{2\pi}\,C_{\gamma\gamma}(\mu_w)\,\ln^2\frac{\mu_w^2}{\mu_0^2} \right] , \\
   \big[ C_{ee}^{V,LR}(\mu_0) \big]_{2\ell\ell2}^{\rm ALP}
   &\approx \big[ C_{ee}^{V,LR}(\mu_w) \big]_{2\ell\ell2}^{\rm ALP}
    - \frac{1}{(4\pi f)^2}\,\frac{32\pi\spac\alpha}{3}\,C_{\gamma\gamma}^2(\mu_w)\,\delta_{2\ell}\,
    \ln\frac{\mu_w^2}{\mu_0^2} \,.
\end{aligned}
\end{equation}

When these results are inserted into \eqref{eq:amumaster} and \eqref{eq:amuALP}, we obtain for the ALP-induced new-physics contribution to the anomalous magnetic moment of the muon in the leading-logarithmic approximation 
\begin{equation}\label{meisterwerk}
\begin{aligned}
   \left[ a_\mu \right]^{\rm ALP}
   &=  - \frac{4 m_\mu}{e}\,{\Re e}\spac\big[ C_{e\gamma}(\mu_w) \big]_{22}^{\rm ALP}
    - m_\mu \sum_{\ell=e,\mu,\tau} \frac{m_\ell}{4\pi^2}\,{\Re e}\spac
    \big[ C_{ee}^{V,LR}(\mu_w) \big]_{2\ell\ell2}^{\rm ALP} \\[-2mm]
   &\quad + \frac{m_\mu^2}{(4\pi f)^2}\,\Bigg\{ - c_{\mu\mu}^2(\mu_w)\,h_1(x_\mu)
    - 8\spac c_{\mu\mu}(\mu_w)\,C_{\gamma\gamma}(\mu_w) 
    \left[ \ln\frac{\mu_w^2}{m_\mu^2} + \delta_2 + 3 - h_2(x_\mu) \right] \\
   &\qquad - \frac{2\spac\alpha(\mu_w)}{\pi}\,c_{\mu\mu}(\mu_w) 
    \sum_{f\ne t}\spac N_c^f\spac Q_f^2\,c_{ff}(\mu_w) \int_0^1\!dz\,F\big(z(1-z)\spac x_f,x_\mu\big) \\
   &\qquad + \frac{12\spac\alpha(\mu_w)}{\pi}\,C_{\gamma\gamma}^2(\mu_w)\, 
    \bigg[ \ln^2\frac{\mu_w^2}{m_\mu^2} 
    + \left( 2\spac\delta_2 + \frac{56}{9} - 2\spac h_2(x_\mu) + \frac{x_\mu}{3} \right) 
    \ln\frac{\mu_w^2}{\mu_0^2} - \ln^2\frac{\mu_0^2}{m_\mu^2} \\
   &\hspace{4.77cm} + \frac{1}{12}\,\Delta_{\rm LbL}(\mu_0,m_a,m_\mu) \bigg] \\[-2mm]
   &\qquad + \frac{m_\tau}{m_\mu}\,{\Re e}\spac\big( [k_e]_{23}\spac[k_E]_{23}^* \big)\,h_3(x_\tau)
    - \frac12\spac\big( |[k_e]_{12}|^2 + |[k_E]_{12}|^2 \big)\,h_4(x_\mu) \Bigg\} \,,
\end{aligned}
\end{equation}
where we consistently neglect higher-order loop corrections to each combination of ALP couplings. This formula is one of the main results of our paper. As explained earlier, it contains the contributions proportional to $c_{\mu\mu}^2$ and $c_{\mu\mu}\,C_{\gamma\gamma}$ at one-loop order and the contributions proportional to $C_{\gamma\gamma}^2$ and $c_{\mu\mu}\,c_{ff}$ at two-loop order, which in each case is the order at which the corresponding terms arise for the first time. Note that the dependence on the low-energy scale $\mu_0$ has cancelled out, as it should. For the terms proportional to $C_{\gamma\gamma}^2$, we conclude that the combination
\begin{equation}\label{eq:49}
   \frac{1}{12}\,\Delta_{\rm LbL}(\mu_0,m_a,m_\mu) - \ln^2\frac{\mu_0^2}{m_\mu^2} 
    - \left( 2\spac\delta_2 + \frac{56}{9} - 2\spac h_2(x_\mu) + \frac{x_\mu}{3} \right) \ln\frac{\mu_0^2}{m_\mu^2}
\end{equation}
must be independent of $\mu_0$. This condition fixes the scale dependence of the quantity $\Delta_{\rm LbL}$. 

The large single logarithm in the term proportional to $c_{\mu\mu}\,C_{\gamma\gamma}$ in the second line of \eqref{meisterwerk} and the large double logarithm in the term proportional to $C_{\gamma\gamma}^2$ in the third line agree with the findings of \cite{Marciano:2016yhf}.\footnote{The ALP couplings used in this paper are related to our couplings by $g_{a\gamma\gamma}=4\spac C_{\gamma\gamma}/f$ and $y_{a\mu}=-m_\mu\spac c_{\mu\mu}/f$.} 
The large single logarithm proportional to $C_{\gamma\gamma}^2$ is a new result of the present work. The authors of \cite{Marciano:2016yhf} have estimated the coefficient of the single logarithm from the ALP-induced vacuum polarization to be $\frac{2}{27}$ in units where the double logarithm has a coefficient~1. The coefficient we obtain is much larger in magnitude (it is $-\frac{16}{9}$ for $m_a/m_\mu\to 0$). Nevertheless, we find that the double-logarithmic contribution still yields the dominant effect.

\subsection{Matching corrections at the electroweak scale}

In order to relate the master formula \eqref{meisterwerk} to the ALP couplings in an effective theory above the electroweak scale -- the SMEFT\,+\,ALP effective theory considered in \cite{Galda:2021hbr} -- it is necessary to express the ALP couplings and LEFT Wilson coefficients entering this result in terms of ALP couplings and SMEFT Wilson coefficients defined in the high-energy theory, which still contains the heavy $W^\pm$, $Z$ and Higgs bosons as well as the top quark. The one-loop matching corrections to the ALP couplings at the scale $\mu_w$ have been calculated in \cite{Bauer:2020jbp}. For the case of the ALP--photon coupling, the relevant relation reads 
\begin{equation}\label{eq:Cggmatch}
\begin{aligned}
   C_{\gamma\gamma}(\mu_w^-) 
   &= C_{\gamma\gamma}(\mu_w^+) - \frac{\alpha(\mu_w)}{4\pi}\,N_c\spac Q_u^2\,c_{tt}(\mu_w^+) \\
   &= c_w^2\,C_{BB}(\mu_w^+) + s_w^2\,C_{WW}(\mu_w^+) - \frac{\alpha(\mu_w)}{3\pi}\,c_{tt}(\mu_w^+) \,,
\end{aligned}
\end{equation}
where the notation $\mu_w^+$ and $\mu_w^-$ means scales infinitesimally above or below the scale $\mu_w$, respectively, and in the second step we have expressed $C_{\gamma\gamma}$ in the high-energy effective theory in terms of the ALP couplings to the $U(1)_Y$ and $SU(2)_L$ gauge bosons, as defined in \cite{Galda:2021hbr}. In this step the electroweak mixing angle (renormalized at the scale $\mu_w$) appears, and we use the short-hand notation
\begin{equation}
   c_w^2\equiv\cos^2\theta_w = \frac{m_W^2}{m_Z^2} \,, \qquad
   s_w^2\equiv\sin^2\theta_w \,.
\end{equation}
The ALP--photon coupling in the higher-energy theory is defined as in \eqref{eq:tildecVV} and \eqref{eq:ALPbosonnew} but with the sum extending over {\em all\,} SM fermions, including the top quark. The above matching relation simply removes the top-quark contribution in the low-energy theory. The original ALP--photon coupling $c_{\gamma\gamma}$ in \eqref{LlowE} does not receive a matching corrections at the electroweak scale. Note that the matching contribution to the QED coupling $\alpha(\mu_w)$ induced by the shift in $C_{\gamma\gamma}$, cf.~\eqref{eq:45}, is of second order in $\alpha$ and hence can be neglected to the order we are working.

The one-loop matching relation for the ALP--muon coupling is more complicated. From \cite{Bauer:2020jbp} we obtain
\begin{equation}
\begin{aligned}
   c_{\mu\mu}(\mu_w^-) 
   &= c_{\mu\mu}(\mu_w^+) + \frac{3\spac\alpha_t(\mu_w)}{4\pi}\,c_{tt}(\mu_w^+)\,\ln\frac{\mu_w^2}{m_t^2} \\
   &\quad - \frac{\alpha_1(\mu_w)}{4\pi}\,C_{BB}(\mu_w^+) \left[
    \left( \frac{15}{2} - 12\spac c_w^4 \right) \left( \ln\frac{\mu_w^2}{m_Z^2} + \delta_1 + \frac12 \right)
    - 6\spac c_w^2 \left( 1 - 4\spac s_w^2 \right) \right] \\
   &\quad - \frac{\alpha_2(\mu_w)}{4\pi}\,C_{WW}(\mu_w^+) \left[
    \left( \frac92 - 12\spac s_w^4 \right) \left( \ln\frac{\mu_w^2}{m_Z^2} + \delta_1 + \frac12 \right)
    - 3\spac\ln c_w^2 + 6\spac s_w^2 \left( 1 - 4\spac s_w^2 \right) \right] ,
\end{aligned}
\end{equation}
where $\alpha_t\equiv y_t^2/(4\pi)$ and $m_t\equiv m_t(m_t)$. The quantity $\delta_1$ is another scheme-dependent constant related to the treatment of the Levi--Civita symbol in dimensional regularization. One finds $\delta_1=-\frac{11}{3}$ in a scheme where $\epsilon^{\mu\nu\alpha\beta}$ is treated as a $d$-dimensional object (our default choice), and $\delta_1=0$ in a scheme where it is treated as a four-dimensional quantity \cite{Bauer:2017ris}.

\begin{figure}
\centering
\includegraphics[scale=1.25]{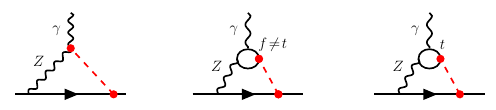} \\[2mm]
\hspace{1.34cm}\includegraphics[scale=1.25]{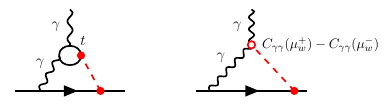}
\caption{Representative Feynman diagrams contributing to the ALP-induced matching corrections to the LEFT dipole coefficient $C_{e\gamma}$ at the electroweak scale. In the second and third diagram in the first row the $Z$ boson can also be replaced by the corresponding Goldstone boson (not shown). The open red circle in the last diagram accounts for the matching contribution obtained from the fact that the ALP--photon coupling $C_{\gamma\gamma}$ differs in the effective theories above and below the electroweak scale by an amount proportional to $c_{tt}$, see \eqref{eq:Cggmatch}.}
\label{fig:amudiags_match}
\end{figure}

\begin{figure}
\centering
\includegraphics[scale=1.25]{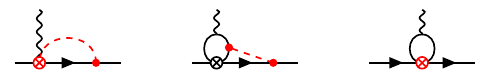}
\caption{Low-energy subgraphs for the three diagrams in the first row of Figure~\ref{fig:amudiags_match} (see the main text for more details). These contributions are power suppressed and do not contribute to the matching.}
\label{fig:subgraphs}
\end{figure}

The LEFT dipole coefficient $C_{e\gamma}$ also receives ALP-induced matching contributions at the electroweak scale, which result from Feynman diagrams with a heavy $Z$ boson and/or top quark in the loop, as shown in Figure~\ref{fig:amudiags_match}.\footnote{There also exist diagrams in which the fermion in the loop is replaced by a $W$ boson. These graphs yield two-loop contributions proportional to $c_{\mu\mu}\spac C_{WW}$, which we neglect for consistency.} 
Analogous diagrams in which the ALP and gauge-boson lines are crossed are not shown. The three $Z$-boson graphs in the first row are proportional to $(g_L^e+g_R^e)=(-\frac12+2\spac s_w^2)\approx -0.04$. They should be evaluated in a general $R_\xi$ gauge, in which case the $Z$ boson can also be replaced by the corresponding Goldstone boson. The result for the first diagram can be inferred from an expression given in \cite{Bauer:2017ris}. The remaining two diagrams have not yet been calculated. We find that the second diagram vanishes (up to a power-suppressed contribution of $\mathcal{O}(m_f^2/m_Z^2)$, which must be dropped for consistency). The third diagram gives a contribution proportional to $c_{\mu\mu}\,c_{tt}$, which is independent of the scale $\mu_w$ but dependent on the ratio $m_t^2/m_Z^2$. It is a reasonable approximation to neglect this unknown contribution, because it is much smaller than other terms involving the same ALP couplings in the result \eqref{eq:Cegammamatch} below. We have checked that the graphs in the first row do not contain any low-energy contributions that would need to be subtracted in the matching. Shrinking the heavy-particle propagators to a point and assuming that at least one of the remaining loop momenta is soft leads to the low-energy subgraphs shown in Figure~\ref{fig:subgraphs}. The effective vertex marked by a red ${\color{red}\otimes}$ symbol in the first subgraph corresponds to the dimension-7 operator $(\partial_\mu a)\,\bar\mu\spac\gamma_\nu\spac F^{\mu\nu}\mu$ in the effective LEFT\,+\,ALP theory, whose coefficient is proportional to $1/(f\spac v^2)$. The dimension-six four-lepton LEFT operator denoted by the black $\otimes$ symbol in the second subgraph has a coefficient of order $1/v^2$. The dimension-7 four-lepton operator denoted by the red ${\color{red}\otimes}$ symbol in the third subgraph has a coefficient of order $1/(f^2\spac v)$, where we include the factor $m_\mu$ from the ALP--muon coupling in the definition of the operator. In all cases, one finds that these low-energy contributions are power suppressed $\sim m_{\mu,a}^2/v^2$ and hence have no effect on the matching calculation. The three diagrams shown in the second row of Figure~\ref{fig:amudiags_match} correspond to the Barr--Zee contribution of the top quark (first graph) as well as a matching contribution arising from the fact that the ALP--photon coupling $C_{\gamma\gamma}$ in the effective theories just above and below the electroweak scale differs by a contribution involving the top quark. Both diagrams contain a non-zero low-energy contribution sensitive to the masses of the ALP and the muon, but in their sum this contribution cancels out.

Using results from \cite{Bauer:2017ris,Buen-Abad:2021fwq,Bauer:2021mvw,Jenkins:2017jig} along with our own calculations, we find that the complete electroweak matching contribution to $C_{e\gamma}$ can be written as
\begin{equation}\label{eq:Cegammamatch}
\begin{aligned}
   \big[ C_{e\gamma}(\mu_w^-) \big]_{22}^{\rm ALP}
   &= \frac{v}{\sqrt2} \left( c_w\spac\big[ C_{eB}(\mu_w^+) \big]_{22}^{\rm ALP} 
    - s_w\spac\big[ C_{eW}(\mu_w^+) \big]_{22}^{\rm ALP} \right) \\
   &\quad - \frac{e\spac m_\mu}{2\spac(4\pi f)^2} \left( 1 - 4\spac s_w^2 \right) c_{\mu\mu}(\mu_w^+)\,
    \big[ C_{BB}(\mu_w^+) - C_{WW}(\mu_w^+) \big] \left( \ln\frac{\mu_w^2}{m_Z^2} + \delta_2 + \frac32 \right) \\
   &\quad + \frac{e\spac m_\mu}{(4\pi f)^2}\,\frac{2\spac\alpha(\mu_w)}{3\pi}\,
    c_{\mu\mu}(\mu_w^+)\,c_{tt}(\mu_w^+) \left[ 
    \left( \ln\frac{\mu_w^2}{m_t^2} + \delta_2 - \frac12 \right) + (1-4\spac s_w^2)\spac(\dots) \right] ,
\end{aligned}
\end{equation}
where the dots refer to the unknown contribution from the third diagram in Figure~\ref{fig:amudiags_match}.

For the LEFT coefficient $C_{ee}^{V,LR}$, the matching relation is of the form \cite{Jenkins:2017jig}
\begin{equation}\label{eq:Ceematch}
   \big[ C_{ee}^{V,LR}(\mu_w^-) \big]_{2\ell\ell2}^{\rm ALP}
   = \big[ C_{le}(\mu_w^+) \big]_{2\ell\ell2}^{\rm ALP} + \text{loop corrections} \,,
\end{equation}
where the relevant one-loop contributions arising from the contact interactions contained in diagrams of the form 
\begin{figure}[h]
\centering
\includegraphics[scale=1.25]{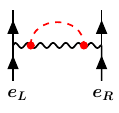}
\end{figure}

\noindent
have not yet been calculated. These effects are quadratic in the ALP--boson couplings $C_{BB}$, $C_{WW}$ and the gauge couplings $g_1$, $g_2$.

Using the matching conditions \eqref{eq:Cggmatch}\spac--\spac\eqref{eq:Ceematch}, the master formula \eqref{meisterwerk} can be reexpressed in terms of the ALP--fermion couplings $c_{\mu\mu}$, $c_{tt}$ and the ALP--boson couplings $C_{BB}$, $C_{WW}$ defined above the electroweak scale $\mu_w$, as well as the ALP contributions to the SMEFT Wilson coefficients $C_{eB}$, $C_{eW}$ and $C_{le}$, all defined at the electroweak scale.

\subsection{RG evolution above the electroweak scale}

The RG evolution equations in the SMEFT\,+\,ALP effective theory are complicated and are best solved in numerical form. A {\tt Mathematica} notebook performing this task, which is based on an implementation of the ALP source terms in {\tt DsixTools~2.0} \cite{Celis:2017hod,Fuentes-Martin:2020zaz}, has been made available in \cite{Biekotter:2023mpd}. The presence of the ALP couplings to top quarks and gluons introduces strong-coupling effects governed by $\alpha_s$ and $\alpha_t$ in the evolution equations. Approximate (but accurate) numerical solutions for the ALP couplings have been derived in \cite{Bauer:2020jbp}. For the relevant couplings to muons and top quarks, one obtains at leading order in RG-improved perturbation theory 
\begin{equation}\label{eq:UVrunning}
\begin{aligned}
   c_{\mu\mu}(\mu_w^+) 
   &\approx c_{\mu\mu}(\Lambda) + \hat I_t(\mu_w,\Lambda)\,c_{tt}(\Lambda)
    - \left[ \frac{15\spac\alpha_1}{8\pi}\,C_{BB}(\Lambda) 
    + \frac{9\spac\alpha_2}{8\pi}\,C_{WW}(\Lambda) \right] \ln\frac{\Lambda^2}{\mu_w^2} \,, \\
   c_{tt}(\mu_w^+) 
   &\approx \left[ 1 - \frac32\,\hat I_t(\mu_w,\Lambda) \right] c_{tt}(\Lambda) 
    - \frac{16}{7} \left[ \frac{\alpha_s(\mu_w)}{\alpha_s(\Lambda)} - 1 \right] C_{GG}(\Lambda) \\
   &\quad - \left[ \frac{17\spac\alpha_1}{24\pi}\,C_{BB}(\Lambda) + \frac{9\spac\alpha_2}{8\pi}\,C_{WW}(\Lambda) \right] 
    \ln\frac{\Lambda^2}{\mu_w^2} \,,
\end{aligned}
\end{equation}
where we have used that $\beta_0^{\rm QCD}=7$ in six-flavor QCD, and \cite{Bauer:2020jbp}\footnote{In this reference the notation $I_t(\mu_w,\Lambda)=-\hat I_t(\mu_w,\Lambda)\,c_{tt}(\Lambda)$ is used.}
\begin{equation}\label{eq:Ihatt}
   \hat I_t(\mu_w,\Lambda)
   = \int_{\mu_w}^\Lambda\!\frac{d\mu}{\mu}\,\frac{3\spac\alpha_t(\mu)}{2\pi}\,\frac{c_{tt}(\mu)}{c_{tt}(\Lambda)}
   = \frac{3\spac\alpha_t(\mu_w)}{\alpha_s(\mu_w)} \left[ 1 
    - \left( \frac{\alpha_s(\Lambda)}{\alpha_s(\mu_w)} \right)^{\!\frac17} \right] ,
\end{equation}
where the result for the integral holds at leading order in RG-improved perturbation theory. In \eqref{eq:UVrunning} we have neglected the scale dependence of the electroweak gauge couplings $\alpha_1$ and $\alpha_2$ as well as of the ALP--boson couplings $C_{BB}$ and $C_{WW}$, which is a good approximation unless $\Lambda=4\pi f$ is exceedingly high above the scale $\mu_w$. In the above expressions the large logarithms associated with powers of $\alpha_s$ and $\alpha_t$ at higher orders in perturbation theory have been resummed to all orders, and one should consider the integral $\hat I_t$ and the factor multiplying $C_{GG}$ as $\mathcal{O}(1)$ quantities.

Our next task is to obtain the relevant SMEFT coefficients by integrating their RG equations including the relevant ALP source terms. We will do this in a simple approximation, by retaining only the leading large logarithms in the scale ratio $\Lambda/\mu_w$. For the ALP coupling to muons, we then obtain 
\begin{equation}\label{eq:ALPmuonlinearized}
\begin{aligned}
   c_{\mu\mu}(\mu_w^+) 
   &\approx c_{\mu\mu}(\Lambda) + \left[ \frac{3\spac\alpha_t}{4\pi}\,c_{tt}(\Lambda)
    - \frac{15\spac\alpha_1}{8\pi}\,C_{BB}(\Lambda) - \frac{9\spac\alpha_2}{8\pi}\,C_{WW}(\Lambda) \right] 
    \ln\frac{\Lambda^2}{\mu_w^2} \,.
\end{aligned}
\end{equation}
This result is needed in the source terms for the SMEFT dipole coefficients $C_{eB}$ and $C_{eW}$, which enter in \eqref{eq:Cegammamatch} \cite{Galda:2021hbr}. Under one-loop scale evolution \cite{Jenkins:2013wua,Alonso:2013hga}, these coefficients mix with each other and with the coefficients $C_{HB}$, $C_{HW}$, $C_{HWB}$, for which there exist non-vanishing ALP source terms, as well as with $C_{H\widetilde B}$, $C_{H\widetilde W}$, $C_{H\widetilde WB}$ and $C_{lequ}^{(3)}$, for which the ALP source terms vanish. Integrating the RG equations for the former coefficients from the high scale $\Lambda=4\pi f$ down to lower energies, we find in leading logarithmic approximation 
\begin{align}
   \big[ C_{HB}(\mu) \big]^{\rm ALP}
   &\approx \big[ C_{HB}(\Lambda) \big]^{\rm ALP}
    + \frac{4\pi\spac\alpha_1}{(4\pi f)^2}\,C_{BB}^2(\Lambda)\,\ln\frac{\Lambda^2}{\mu^2} \,, \notag\\
   \big[ C_{HW}(\mu) \big]^{\rm ALP}
   &\approx \big[ C_{HW}(\Lambda) \big]^{\rm ALP}
    + \frac{4\pi\spac\alpha_2}{(4\pi f)^2}\,C_{WW}^2(\Lambda)\,\ln\frac{\Lambda^2}{\mu^2} \,, \\
   \big[ C_{HWB}(\mu) \big]^{\rm ALP}
   &\approx \big[ C_{HWB}(\Lambda) \big]^{\rm ALP}
    + \frac{8\pi\spac\sqrt{\alpha_1\spac\alpha_2}}{(4\pi f)^2}\,C_{BB}(\Lambda)\,C_{WW}(\Lambda)\,
    \ln\frac{\Lambda^2}{\mu^2} \,. \notag
\end{align}
Using the above relations, it is straightforward to integrate the RG equations for the dipole coefficients in the leading logarithmic approximation. We obtain 
\begin{equation}
\begin{aligned}
   \big[ C_{eB}(\mu_w^+) \big]_{22}^{\rm ALP} 
   &\approx \big[ C_{eB}(\Lambda) \big]_{22}^{\rm ALP}
    + \frac{y_\mu}{16\pi^2} \left[ \frac{3\spac g_1}{2}\,\big[ C_{HB}(\Lambda) \big]^{\rm ALP}
    - \frac{3\spac g_2}{4}\,\big[ C_{HWB}(\Lambda) \big]^{\rm ALP} \right] \ln\frac{\Lambda^2}{\mu_w^2} \\
   &\quad + \frac{g_1\spac y_\mu}{(4\pi f)^2}\,C_{BB}(\Lambda)\,\bigg\{
    \frac32\,c_{\mu\mu}(\Lambda)\,\ln\frac{\Lambda^2}{\mu_w^2} \\
   &\qquad + \left[ \frac{9\spac\alpha_t}{16\pi}\,c_{tt}(\Lambda)
    - \frac{39\spac\alpha_1}{32\pi}\,C_{BB}(\Lambda) - \frac{33\spac\alpha_2}{32\pi}\,C_{WW}(\Lambda)
    \right] \ln^2\frac{\Lambda^2}{\mu_w^2} \bigg\} \,, \\
\end{aligned}
\end{equation}
and
\begin{equation}
\begin{aligned}
   \big[ C_{eW}(\mu_w^+) \big]_{22}^{\rm ALP}
   &\approx \big[ C_{eW}(\Lambda) \big]_{22}^{\rm ALP}
    + \frac{y_\mu}{16\pi^2} \left[ \frac{g_2}{2}\,\big[ C_{HW}(\Lambda) \big]^{\rm ALP}
    - \frac{3\spac g_1}{4}\,\big[ C_{HWB}(\Lambda) \big]^{\rm ALP} \right] \ln\frac{\Lambda^2}{\mu_w^2} \\
   &\quad + \frac{g_2\,y_\mu}{(4\pi f)^2}\,C_{WW}(\Lambda)\,\bigg\{
    - \frac12\,c_{\mu\mu}(\Lambda)\,\ln\frac{\Lambda^2}{\mu_w^2} \\
   &\qquad + \left[ - \frac{3\spac\alpha_t}{16\pi}\,c_{tt}(\Lambda)
    + \frac{21\spac\alpha_1}{32\pi}\,C_{BB}(\Lambda) + \frac{7\spac\alpha_2}{32\pi}\,C_{WW}(\Lambda) \right]
    \ln^2\frac{\Lambda^2}{\mu_w^2} \bigg\} \,.
\end{aligned}
\end{equation}
Finally, for the SMEFT coefficient $C_{le}$ entering in \eqref{eq:Ceematch}, it is sufficient to integrate the ALP source term directly, which leads to 
\begin{equation}
   \big[ C_{le}(\mu_w^+) \big]_{2\ell\ell2}^{\rm ALP}
   \approx \big[ C_{le}(\Lambda) \big]_{2\ell\ell2}^{\rm ALP}
    - \frac{1}{(4\pi f)^2}\,\frac{16\pi\spac\alpha_1}{3}\,C_{BB}^2(\Lambda)\,\delta_{2\ell}\,
    \ln\frac{\Lambda^2}{\mu_w^2} \,.
\end{equation}

\subsection{Models with loop-suppressed ALP--boson couplings}

In many realistic ALP models, the ALP--boson couplings are loop suppressed, as indicated in \eqref{LlowE} and \eqref{LlowEalt}, in which case the parameters $C_{BB}$ and $C_{WW}$ are likely to be much smaller than the ALP--fermion couplings $c_{\mu\mu}$ and $c_{tt}$. In such scenarios, the master formula \eqref{meisterwerk} can be simplified and rewritten in the form 
\begin{equation}
\begin{aligned}
   \left[ a_\mu \right]^{\rm ALP}
   &=  - \frac{4 m_\mu}{e}\,{\Re e}\spac\big[ C_{e\gamma}(\mu_w^-) \big]_{22}^{\rm ALP} \\
   &\quad - \frac{m_\mu^2}{(4\pi f)^2}\,c_{\mu\mu}(\mu_w)\,\Bigg\{ 
    c_{\mu\mu}(\mu_w^-)\,h_1(x_\mu)
    + \frac{2\spac\alpha(\mu_w)}{\pi}\,\tilde c_{\gamma\gamma}(\mu_w^-) 
    \left[ \ln\frac{\mu_w^2}{m_\mu^2} + \delta_2 + 3 - h_2(x_\mu) \right] \\
   &\hspace{4.0cm} + \frac{2\spac\alpha(\mu_w)}{\pi} 
    \sum_{f\ne t}\spac N_c^f\spac Q_f^2\,c_{ff}(\mu_w^-) \int_0^1\!dz\,F\big(z(1-z)\spac x_f,x_\mu\big) \Bigg\} \,,
\end{aligned}
\end{equation}
with $\tilde c_{\gamma\gamma}$ defined in \eqref{eq:tildecVV}. The term proportional to $\tilde c_{\gamma\gamma}^2$ is now of four-loop order and can safely be neglected, with the leading contribution given by
\begin{equation}
   \frac{m_\mu^2}{(4\pi f)^2}\,\frac34 \left[ \frac{\alpha(\mu_w)}{\pi} \right]^3 
    \tilde c_{\gamma\gamma}^2(\mu_w)\,\ln^2\frac{m_Z^2}{m_\mu^2} 
   \approx 1.5\cdot 10^{-14}\,\spac\tilde c_{\gamma\gamma}^2(\mu_w) \left[ \frac{100\,\text{GeV}}{f} \right]^2 .
\end{equation}
There is no ALP source term for the Wilson coefficient $C_{ee}^{V,LR}$ in this case, and hence we have dropped the corresponding contribution. For simplicity, we have also omitted the contributions involving flavor-changing ALP couplings. They can be added back if desired. 

In this simpler context, it is straightforward to implement the relations which accomplish the extrapolation to the new-physics scale $\Lambda=4\pi f$ at leading logarithmic approximation. We obtain the master formula 
\begin{equation}\label{meisterwerk2}
\begin{aligned}
   \left[ a_\mu \right]^{\rm ALP}
   &=  - \frac{4 m_\mu}{e}\,\frac{v}{\sqrt2}\,{\Re e}\!\left( c_w\spac\big[ C_{eB}(\Lambda) \big]_{22}^{\rm ALP} 
    - s_w\spac\big[ C_{eW}(\Lambda) \big]_{22}^{\rm ALP} \right) \\[-1mm]
   &\quad - \frac{m_\mu^2}{(4\pi f)^2}\,
    \Bigg\{ \Big[ c_{\mu\mu}(\Lambda) + \hat I_t(m_t,\Lambda)\,c_{tt}(\Lambda) \Big]^2\,h_1(x_\mu) \\
   &\hspace{2.5cm} + \frac{2\spac\alpha}{\pi}\,\Big[ c_{\mu\mu}(\Lambda) 
    + \hat I_t(m_t,\Lambda)\,c_{tt}(\Lambda) \Big] \\
   &\hspace{2.5cm} \times \Bigg[ \left( \frac{3\spac\tilde c_{BB}(\Lambda)}{4\spac c_w^2} 
    + \frac{\tilde c_{WW}(\Lambda)}{4\spac s_w^2} \right) 
    \left( \ln\frac{\Lambda^2}{m_Z^2} + \delta_2 + \frac32 \right) \\
   &\hspace{3.22cm} + \Big( \tilde c_{BB}(\Lambda) + \tilde c_{WW}(\Lambda) \Big) 
    \left[ \ln\frac{m_Z^2}{m_\mu^2} + \frac32 - h_2(x_\mu) \right] \\
   &\hspace{3.22cm} + \sum_f\spac N_c^f\spac Q_f^2\,c_{ff}(\Lambda)
    \int_0^1\!dz\,F\big(z(1-z)\spac x_f,x_\mu\big) \Bigg] \Bigg\} \,,
\end{aligned}
\end{equation}
where the integral $\hat I_t$ has been given in \eqref{eq:Ihatt}, and as before we neglect higher-order loop corrections to each combination of ALP couplings. Note that the electroweak matching scale $\mu_w$ has dropped out, as it should. The QED coupling $\alpha$ and the electroweak mixing angle should be evaluated at the scale $\mu_w\sim m_Z$. The sum in the last line now extends over {\em all\/} quark flavors, and we have used that for the top quark
\begin{equation}
   \int_0^1\!dz\,F\big(z(1-z)\spac x_t,x_\mu\big)
   = - \ln\frac{m_t^2}{m_\mu^2} + h_2(x_\mu) - \frac72 
\end{equation}
up to power corrections of $\mathcal{O}(m_{\mu,a}^2/m_t^2)$. We stress again that the large logarithms $\ln(\Lambda^2/m_Z^2)$, $\ln(m_Z^2/m_\mu^2)$ and $\ln(m_t^2/m_\mu^2)$ contained in the above results could be straightforwardly resummed to all orders of perturbation theory by solving the RG equations in the LEFT and the SMEFT numerically, using tools such as those provided in \cite{Biekotter:2023mpd}.

Relation \eqref{meisterwerk2} is our final result for the ALP-induced contribution to the anomalous magnetic moment of the muon in models with loop-suppressed ALP--boson couplings. Note the remarkable fact that RG evolution effects shift the ALP--muon coupling by an amount proportional to the ALP--top coupling, and hence a non-zero contribution to $a_\mu$ can be generated even if the ALP--muon coupling is suppressed (or absent) in the UV theory. It is well known that the contribution proportional to $c_{\mu\mu}^2$ is negative and hence has the wrong sign to explain the current tension seen between the experimental value of $a_\mu$ and its theoretical prediction \cite{Aoyama:2020ynm}. In order to obtain an overall positive ALP contribution, the remaining terms must yield significant contributions of the correct sign. 

\begin{table}
\centering
\begin{tabular}{c|ccc|ccc}
\toprule
\rowcolor{LightGray}
 & \multicolumn{3}{c|}{$f=100$\,GeV, $\Lambda=1.26$\,TeV} & \multicolumn{3}{c}{$f=300$\,GeV, $\Lambda=3.77$\,TeV} \\ 
\rowcolor{VeryLightGray}
ALP couplings & $m_a=0$ & $0.3$\,GeV & $1$\,GeV &$m_a=0$ & $0.3$\,GeV & $1$\,GeV \\ 
\midrule
$c_{\mu\mu}^2(\Lambda)$ & $-668.1$ & $-166.8$ & $-32.5$
 & $-74.2$ & $-18.5$ & $-3.6$ \\
\rowcolor{VeryLightGray}
$c_{\mu\mu}(\Lambda)\,\tilde c_{BB}(\Lambda)$ & $-62.2$ & $-52.0$ & $-44.4$
 & $-7.7$ & $-6.6$ & $-5.8$ \\
$c_{\mu\mu}(\Lambda)\,\tilde c_{WW}(\Lambda)$ & $-63.6$ & $-53.4$ & $-45.8$
 & $-8.0$ & $-6.9$ & $-6.0$ \\
\rowcolor{VeryLightGray}
$c_{\mu\mu}(\Lambda)\,c_{tt}(\Lambda)$ & $\phantom{-}31.7$ & $\phantom{-}66.9$ & $\phantom{-}64.7$
 & $-0.66$ & $\phantom{-}6.3$ & $\phantom{-}7.0$ \\
$c_{\mu\mu}(\Lambda)\,c_{dd}(\Lambda)$ & $\phantom{-}15.0$ & $\phantom{-}9.0$ & $\phantom{-}5.9$
 & $\phantom{-}1.7$ & $\phantom{-}1.0$ & $\phantom{-}0.65$ \\
\rowcolor{VeryLightGray}
$c_{tt}(\Lambda)\,\tilde c_{BB}(\Lambda)$ & $-3.8$ & $-3.2$ & $-2.7$
 & $-0.69$ & $-0.59$ & $-0.51$ \\
$c_{tt}(\Lambda)\,\tilde c_{WW}(\Lambda)$ & $-3.9$ & $-3.3$ & $-2.8$
 & $-0.71$ & $-0.61$ & $-0.54$ \\
\rowcolor{VeryLightGray}
$c_{tt}^2(\Lambda)$ & $\phantom{-}4.5$ & $\phantom{-}4.7$ & $\phantom{-}4.1$
 & $\phantom{-}0.53$ & $\phantom{-}0.71$ & $\phantom{-}0.65$ \\
$c_{tt}(\Lambda)\,c_{dd}(\Lambda)$ & $\phantom{-}0.92$ & $\phantom{-}0.55$ & $\phantom{-}0.36$
 & $\phantom{-}0.15$ & $\phantom{-}0.089$ & $\phantom{-}0.058$ \\
\bottomrule
\end{tabular}
\caption{\label{tab:coefs}
Contributions to $[a_\mu]^{\rm ALP}$ (in units of $10^{-11}$) proportional to different combinations of ALP couplings at the scale $\Lambda=4\pi f$. We assume flavor-universal ALP--fermion couplings.}
\end{table}

The SM value for the anomalous magnetic moment of the muon is quoted in \cite{Aoyama:2020ynm} as $[a_\mu]^{\rm SM}=116\,591\,810(43)\cdot 10^{-11}$, whereas the present experimental average value is $[a_\mu]^{\rm exp}=116\,592\,061(41)\cdot 10^{-11}$ \cite{Muong-2:2021ojo}. This yields a deviation of $[a_\mu]^{\rm exp}-[a_\mu]^{\rm SM}=(251\pm 59)\cdot 10^{-11}$, corresponding to a $4.2\spac\sigma$ tension. However, there is currently an intense discussion about whether the data-driven estimation of the hadronic contributions to $a_\mu$ used in \cite{Aoyama:2020ynm} is in agreement with recent determinations of these effects in lattice QCD as well as with recent data reported by the CMD-3 collaboration \cite{CMD-3:2023alj}. For example, if the lattice result for the (leading-order) hadronic vacuum polarization reported by the BMW collaboration \cite{Borsanyi:2020mff} is used instead of the data-driven value, the deviation is reduced to about $(107\pm 70)\cdot 10^{-11}$, corresponding to a tension of less than $2\spac\sigma$. In Table~\ref{tab:coefs} we show the ALP-induced contributions to the anomalous magnetic moment of the muon in the form of coefficients multiplying different combinations of ALP couplings, defined at the UV scale $\Lambda=4\pi f$. We assume flavor-universal ALP couplings, so that $c_{ee}=c_{\tau\tau}=c_{\mu\mu}$, $c_{uu}=c_{cc}=c_{tt}$, and $c_{ss}=c_{bb}=c_{dd}$. Note that we omit the UV contribution shown in the first line of \eqref{meisterwerk2}, which is not logarithmically enhanced but needs to be evaluated in the context of a specific ALP model. It is not difficult to generate phenomenologically interesting values of $[a_\mu]^{\rm ALP}$ using reasonable ALP parameters, even in minimal scenarios with only two non-zero couplings. For example, a positive contribution of about $200\cdot 10^{-11}$ is obtained for $f=100$\,GeV, $m_a=300$\,MeV and the ALP couplings $c_{\mu\mu}(\Lambda)=1$ and $\tilde c_{BB}(\Lambda)=-7$, while a contribution of about $100\cdot 10^{-11}$ is found for $f=300$\,GeV, $m_a=1$\,GeV and non-zero couplings $c_{\mu\mu}(\Lambda)=4$ and $\tilde c_{BB}(\Lambda)=-7$. The new-physics scales in the two cases are 1.26\,TeV and 3.77\,TeV, respectively. These are just two out of a large number of possible scenarios. Both sets of parameters are compatible with present bounds from both direct and indirect searches, see e.g.\ the recent discussions in \cite{Bauer:2021mvw,Biekotter:2023mpd}.

\section{Conclusions}
\label{sec:Conclusion}

Axions and axion-like particles (ALPs) are prominent and well-motivated candidates for physics beyond the Standard Model. They may explain the absence of CP-violating effects in the strong interactions (the so-called ``strong CP problem'') and, more generally, they can arise as pseudo Nambu--Goldstone bosons in a large class of extensions of the Standard Model with a broken global $U(1)$ symmetry. In general, ALPs couple to the known particles via dimension-five and yet higher-dimensional operators and hence fall in the class of naturally light and weakly coupled new-physics particles. 

Direct searches for ALPs are being performed using a large variety of experimental probes, ranging from astrophysical measurements to collider searches. Generally, such searches suffer from a strong dependence on the ALP lifetime, its branching ratios for decays into Standard Model particles, and assumptions about the coupling structure of the underlying ALP model. Indirect searches for ALP-induced quantum corrections to precision observables offer an interesting alternative strategy, which alleviates some of these drawbacks. Even if -- as we assume here -- the ALP mass is smaller than the scale of electroweak symmetry breaking, ALPs nevertheless leave an imprint on the effective Wilson coefficients accounting for new-physics effects in the context of the effective field-theory extensions of the Standard Model called SMEFT (above the electroweak scale) and LEFT (below the electroweak scale). The reason is that ALPs can give rise to UV-divergent contributions to Green's functions with external Standard Model particles only. We have first shown this in the context of the SMEFT in \cite{Galda:2021hbr}, where we found that ALPs provide source terms for almost all dimension-6 SMEFT operators.

In the present work, we have extended this analysis to the low-energy effective theory (LEFT), in which the top quark, the $W^\pm$ and $Z$ bosons, and the Higgs boson are integrated out. We have presented the complete set of ALP source terms entering the renormalization-group equations for the LEFT Wilson coefficients. As an important application of this formalism, we have presented a new state-of-the-art calculation of ALP-induced contributions to the anomalous magnetic moment of the muon, which includes all relevant two-loop effects and the UV boundary conditions in the SMEFT. Equation~\eqref{meisterwerk} expresses the result in terms of ALP couplings and LEFT Wilson coefficients defined at the electroweak scale, while in relation~\eqref{meisterwerk2} we show the result expressed in terms of ALP couplings and SMEFT Wilson coefficients defined at the UV scale $\Lambda=4\pi f$. We have presented numerical results for the different ALP-induced contributions to $a_\mu$ in Table~\ref{tab:coefs}, assuming for simplicity flavor-universal ALP--fermion couplings. These results can be applied to a large variety of ALP models.

\subsubsection*{Acknowledgements}

The Feynman diagrams in this paper have been generated using the TikZ-Feynman package \cite{Ellis:2016jkw}. This work has been supported by the Cluster of Excellence Precision Physics, Fundamental Interactions, and Structure of Matter (PRISMA$^+$ EXC 2118/1) funded by the German Research Foundation (DFG) within the German Excellence Strategy (Project ID 39083149).

\begin{appendix}

\section{Dimension-8 four-fermion operators in the LEFT}
\label{app:A}

\renewcommand{\theequation}{A.\arabic{equation}}
\setcounter{equation}{0}

In Section~\ref{subsec:3.3}, we have discussed that four-fermion operators in the LEFT, which originate when the heavy electroweak gauge bosons $W^\pm$ and $Z$ are integrated out, give rise to additional ALP source terms for four-fermion operators, which appear first at dimension-8 order. As an example of such contributions, we consider four-fermion operators with flavor structure $(\bar u\spac u)(\bar e\spac e)$, which arise from $Z$-boson exchange. There are six diagrams to consider, where the ALP is exchanged between any pair of fermions in the effective four-fermion operator. From the $1/\epsilon$ poles of these diagrams, we obtain the following contributions to the source terms: 
\begin{equation}
\begin{aligned}
   \big[ S_{eu}^{V,LR} \big]_{prst}
   &= \frac{4}{v^2}\,\Big( 16\spac\delta_{st}\,g_R^e\spac g_R^u\,[\tilde{\bm{M}}_e\spac\tilde{\bm{M}}_e^\dagger]_{pr} 
    - \delta_{pr}\,g_L^e\spac g_L^u\,[\tilde{\bm{M}}_u^\dagger\spac\tilde{\bm{M}}_u]_{st} \Big) \,, \\
   \big[ S_{ue}^{V,LR} \big]_{prst}
   &= \frac{4}{v^2}\,\Big( 16\spac\delta_{st}\,g_L^e\spac g_L^u\,[\tilde{\bm{M}}_e^\dagger\spac\tilde{\bm{M}}_e]_{pr} 
    - \delta_{pr}\,g_R^e\spac g_R^u\,[\tilde{\bm{M}}_u\spac\tilde{\bm{M}}^\dagger_u ]_{st} \Big) \,, \\
   \big[ S_{eu}^{V,LL} \big]_{prst} 
   &= \frac{4}{v^2}\,\Big( 16\spac\delta_{st}\,g_R^e\spac g_L^u\,[\tilde{\bm{M}}_e\spac\tilde{\bm{M}}^\dagger_e ]_{pr} 
    - \delta_{pr}\,g_L^e\spac g_R^u\,[\tilde{\bm{M}}_u\spac\tilde{\bm{M}}^\dagger_u]_{st} \Big) \,, \\
   \big[ S_{eu}^{V,RR} \big]_{prst}
   &= \frac{4}{v^2}\,\Big( 16\spac\delta_{st}\,g_L^e\spac g_R^u\,[\tilde{\bm{M}}_e^\dagger\spac\tilde{\bm{M}}_e]_{pr}
    - \delta_{pr}\,g_R^e\spac g_L^u\,[\tilde{\bm{M}}^\dagger_u\spac\tilde{\bm{M}}_u]_{st} \Big) \,,
\end{aligned}
\end{equation}
and
\begin{equation}
\begin{aligned}
   \big[ S_{eu}^{S,RL} \big]_{prst} 
   &= \frac{2}{v^2}\,[\tilde{\bm{M}}_e]_{pr}\,[\tilde{\bm{M}}_u^\dagger]_{st} \,, \\
   \big[ S_{eu}^{T,RR} \big]_{prst} 
   &= \frac{2}{v^2}\,(g_R^e + g_L^e)\spac(g_R^u + g_L^u)\,
    [\tilde{\bm{M}}_e]_{pr}\,[\tilde{\bm{M}}_u]_{st} \,, \\
   \big[ S_{eu}^{S,RR} \big]_{prst} 
   &= \frac{2}{v^2}\,[\tilde{\bm{M}}_e]_{pr}\,[\tilde{\bm{M}}_u]_{st} \,, 
\end{aligned}
\end{equation}
where $g_L^e=(-\frac12+s_w^2)$, $g_R^e=s_w^2$ and $g_L^u=(\frac12-\frac23\spac s_w^2)$, $g_R^u=-\frac23\spac s_w^2$ denote the relevant $Z$-boson couplings. The operators corresponding to the source terms in the last equation are not hermitian, and for the hermitian-conjugate operators one needs to exchange $\tilde{\bm{M}}_f\leftrightarrow\tilde{\bm{M}}_f^\dagger$ in the source terms.

\end{appendix}

\end{document}